\documentclass[traditabstract]{aa}
\usepackage{txfonts}
\usepackage{graphicx}

\def\cc{\,{\rm cm^{-3}}}
\def\cm2{\,{\rm cm^{-2}}}
\def\pc2{\,{\rm pc^{2}}}
\def\kms{\,{\rm {km\,s^{-1}}}}
\def\thirco{\,{\rm ^{13}CO}}
\def\h2{\,{\rm H_{2}}}
\def\kkms{\,{\rm {K\,km s^{-1}}}}
\def\co{\,{\rm ^{12}CO}}

\def\ci{\rm [C$\;$I]}
\def\cii{\rm [C$\;$II]}
\def\nii{\rm [N$\;$II]}

\def\hi{\rm H$\;$I}

\def\cp{{\rm C^{+}}}
\def\c{{\rm C^{\circ}}}

\def\coa10{$^{12}$CO($1\rightarrow 0$)}
\def\cob21{$^{12}$CO($2\rightarrow 1$)}
\def\coc32{$^{12}$CO($3\rightarrow 2$)}
\def\cod43{$^{12}$CO($4\rightarrow 3$)}
\def\coe54{$^{12}$CO($5\rightarrow 4$)}
\def\cof76{$^{12}$CO($7\rightarrow 6$)}
\def\13coc32{$^{13}$CO($3\rightarrow 2$)}

\def\hcna10{HCN($1\rightarrow 0$)}
\def\hcnb43{HCN($4\rightarrow 3$)}
\def\car10{CI($^3P_1\rightarrow ^3P_0$)} 
 
\def\carb21{CI($^3P_2\rightarrow ^3P_1)$} 
\def\etal{et al.\ }

\def\psqcm{\ifmmode {\&gt;{\rm cm}^{-2}}\else {cm$^{-2}$}\fi}
\def\pcubcm{\ifmmode {\&gt;{\rm cm}^{-3}}\else {cm$^{-3}$}\fi}
\def\be{\begin{equation}}
\def\ee{\end{equation}}
\def\bea{\begin{eqnarray}}
\def\eea{\end{eqnarray}}
\def\rpcsq{\ifmmode {r_{\rm pc}^2}\else {$r_{\rm pc}^2$}\fi}
\def\rpc{\ifmmode {r_{\rm pc}}\else {$r_{\rm pc}$}\fi}
\def\fabs{\ifmmode {f_{\rm abs}}\else {$f_{\rm abs}$}\fi}
\def\msol{\ifmmode {\&gt;M_\odot}\else {$M_\odot$}\fi}
\def\lsol{\ifmmode {\&gt;L_\odot}\else {$L_\odot$}\fi}
\def\ergsr{\ifmmode {\rm\&gt; erg\;cm^{-2}\;s^{-1}\;sr^{-1}} 
            \else {erg cm$^{-2}$ s$^{-1}$ sr$^{-1}$}\fi}
\def\ergs{\ifmmode {\rm\&gt; erg\;cm^{-2}\;s^{-1}} 
            \else {erg cm$^{-2}$ s$^{-1}$}\fi}
\def\Msun{\rm M_{\odot}}

\def\aua{{\rm A\&A,} }

\def\aar{{\rm A\&AR,} }
\def\apj{{\rm ApJ,} }
\def\aj{{\rm AJ,} }

\def\apjl{{\rm ApJL,} }

\def\mnras{{\rm MNRAS,} }

\begin{document}

\title{The molecular circumnuclear disk (CND) in Centaurus A}

\subtitle{A multi-transition CO and [CI] survey with Herschel, APEX, JCMT, and SEST}

\author{F.P. Israel\inst{1}
 \and   R. G\"usten\inst{2}
 \and   R. Meijerink\inst{3}
 \and   A.F. Loenen\inst{1}
 \and   M.A. Requena-Torres\inst{2} 
 \and   J. Stutzki\inst{4} 
 \and   P. van der Werf\inst{1} 
 \and   A. Harris\inst{5} 
 \and   C. Kramer\inst{6}
 \and   J. Martin-Pintado\inst{7}
 \and   A. Weiss,\inst{2}
        }

\offprints{F.P. Israel}
 
  \institute{Sterrewacht Leiden, Leiden University, P.O. Box 9513,
             2300 RA Leiden, The Netherlands 
  \and       Max-Planck-Institut f\"ur Radioastronomie, 
             Auf dem H\"ugel 69, 53121 Bonn, Germany
  \and       Kapteyn Astronomical Institute, Postbus 800, 9700 AV 
             Groningen, the Netherlands
  \and       I. Physikalisches Institut der Universit\"at zu K\"oln,
             Z\"ulpicher Strasse 77, D-50937 K\"oln, Germany  
  \and       Department of Astronomy, University of Maryland, College Park, 
             MD 20742, USA           
  \and       IRAM, Avenida Divina Pastora, 7, N\'ucleo Central E 18012 Granada, 
             Spain 
  \and       CSIC/INTA, Ctra de Torrej\'on a Ajalvir, km 4, 28850 Torrej\'on de 
             Ardoz, Madrid, Spain 
}

\authorrunning{F.P. Israel \etal }

\titlerunning{Cen A circumnuclear disk}

\date{Received ????; accepted ????}
 
\abstract{This paper presents emission line intensities of CO and C$^{\circ}$
  from the compact circumnuclear disk in the center of NGC~5128
  (Centaurus~A) obtained with the {\it Herschel Space Observatory} in
  the 400-1000 GHz range as well as previously unpublished
  measurements obtained with the ground-based observatories {\it
    SEST}, {\it JCMT} and {\it APEX} in the 90-800 GHz range.  The
  results show that the Cen~A center has an emission ladder of CO
  transitions quite different from those of either star-burst galaxies
  or (Seyfert) AGNs.  In addition, the neutral carbon ([CI]) emission
  lines from the Cen~A center are much stronger relative to the
  adjacent CO lines than in any other galaxy.  The CO surface
  brightness of the compact circumnuclear disk (CND) is significantly
  higher than that of the much more extended thin disk (ETD) in the
  same line of sight.  LVG analysis of the CO line profiles decomposed
  into the constituent contributions show that the ETD is relatively
  cool and of low excitation, wheres the brighter CND is hotter and
  more highly excited.  Our PDR/XDR models suggest that most of the
  CND gas is relatively cool (temperatures 25 K - 80 K) and not very
  dense ($\approx\,300\,\cc$) if it is primarily heated by UV photons.
  A small fraction of the gas in both the CND and the ETD has a much
  higher density (typically $30 000\,\cc$).  A more highly excited,
  high-density phase is present in the CND, either in the form of an
  extreme PDR or more likely in the form of an XDR.  Such a phase does
  not occur in the part of the ETD sampled.  We have determined, for
  the first time, the molecular mass parameters of the CND. The total
  gas mass of the CND is $M_{CND}\,=\,8.4\times10^{7}\,\Msun$,
  uncertain by a factor of two.  The CO-$\h2$ conversion factor
  ($X_{CND}$) is $4\times10^{20}$ (K $\kms$)$^{-1}$ also within a
  factor of two.}  


\maketitle
 
\section{Introduction}

\begin{table}
\scriptsize
\caption[]{Log of Herschel observations}
\begin{center}
\begin{tabular}{ccccr}
\noalign{\smallskip}     
\hline
\noalign{\smallskip}
Instru- & Transi-       & OBSID      & Date  & Integr. \\
ment    & tion          &            & Y-M-D & (sec)  \\
\noalign{\smallskip}     
\hline
\noalign{\smallskip}     
{\it HIFI} & CO $J$=5-4 C    & 1342200931 & 2010-07-19 & 1863 \\
{\it HIFI} & CO $J$=5-4 NW   & 1342200932 & 2010-07-19 &   86 \\
{\it HIFI} & CO $J$=5-4 SE   & 1342200933 & 2010-07-19 &   86 \\
{\it HIFI} & 13CO $J$=5-4 C  & 1342201090 & 2010-07-21 &  679 \\
{\it HIFI} & 13CO $J$=5-4 NW & 1342201093 & 2010-07-21 &  353 \\
{\it HIFI} & 13CO $J$=5-4 SE & 1342201095 & 2010-07-21 &  353 \\
{\it HIFI} & C18O $J$=5-4 C  & 1342201091 & 2010-07-21 &  112 \\
{\it HIFI} & CO $J$=6-5 C    & 1342200982 & 2010-07-19 &   82 \\
{\it HIFI} & CO $J$=6-5 NW   & 1342200983 & 2010-07-19 &   82 \\
{\it HIFI} & CO $J$=6-5 SE   & 1342200984 & 2010-07-19 &   82 \\
{\it HIFI} & CO $J$=7-6 C    & 1342201711 & 2010-07-30 & 1753 \\
{\it HIFI} & CO $J$=7-6 NW   & 1342201714 & 2010-07-30 &   94 \\
{\it HIFI} & CO $J$=7-6 SE   & 1342201715 & 2010-07-30 &   94 \\
{\it HIFI} & CO $J$=8-7 C    & 1342201728 & 2010-07-30 &   92 \\
{\it HIFI} & CO $J$=8-7 NW   & 1342201729 & 2010-07-30 &   92 \\
{\it HIFI} & CO $J$=8-7 SE   & 1342201730 & 2010-07-30 &   92 \\
{\it HIFI} & CO $J$=9-8 C    & 1342200948 & 2010-07-19 & 1896 \\
{\it HIFI} & CO $J$=9-8 NW   & 1342200949 & 2010-07-19 &   70 \\
{\it HIFI} & CO $J$=9-8 SE   & 1342200950 & 2010-07-19 &   70 \\
{\it HIFI} & CO $J$=10-9 C   & 1342201108 & 2010-07-21 &  257 \\
{\it HIFI} & CO $J$=10-9 NW  & 1342201109 & 2010-07-21 &  257 \\
{\it HIFI} & CO $J$=10-9 SE  & 1342201110 & 2010-07-21 &  257 \\
{\it HIFI} & CO $J$=13-12 C  & 1342201775 & 2010-07-30 & 1896 \\
{\it HIFI} & CO $J$=13-12 NW & 1342201776 & 2010-07-30 &   70 \\
{\it HIFI} & CO $J$=13-12 SE & 1342201778 & 2010-07-30 &   70 \\
{\it HIFI} & [CI] $J$=1-0 C  & 1342201089 & 2010-07-21 &  353 \\
{\it HIFI} & [CI] $J$=1-0 NW & 1342201092 & 2010-07-21 &  112 \\
{\it HIFI} & [CI] $J$=1-0 SE & 1342201094 & 2010-07-21 &  112 \\
{\it HIFI} & [CI] $J$=2-1 C  & 1342201712 & 2010-07-30 &   88 \\
{\it HIFI} & [CI] $J$=2-1 NW & 1342201713 & 2010-07-30 &   88 \\
{\it HIFI} & [CI] $J$=2-1 SE & 1342201716 & 2010-07-19 & 1884 \\
{\it HIFI} & [CII] C         & 1342213717 & 2011-02-04 & 8970 \\
{\it HIFI} & [CII] NW        & 1342201643 & 2010-07-28 &  936 \\
{\it HIFI} & [CII] SE        & 1342201644 & 2010-07-28 &  936 \\
{\it HIFI} & [NII]           & 1342201778 & 2010-07-31 &  833 \\ 
{\it SPIRE}& SSPEC           & 1342204037 & 2010-08-23 & 5041 \\
\noalign{\smallskip}     
\hline
\noalign{\smallskip}
\end{tabular}
\end{center}
\label{herschellog}
\end{table}

\nobreak Disks of dense dust and gas deeply embedded in the stellar
body of a giant elliptical galaxy are the more easily identified
remnants of smaller gas-rich galaxies that have fallen in. They are a
transient phenomenon: on varying time-scales, the gas will be consumed
by accretion onto a central black hole, expulsion by jets emanating
from the nucleus, and by the in-situ formation of new stars
rejuvenating the host galaxy stellar population. All three processes
are currently active in NGC~5128, the host of the Fanaroff-Riley class
I (FR I) radio source Centaurus A (Cen A).

\nobreak The properties of these embedded disks are probably different
from those of the interstellar medium (ISM) in spiral or star-burst
galaxies (see e.g. the review by Henkel $\&$ Wiklind 1997).
Kiloparsec-sized embedded thin disks are directly exposed to the
intense but UV-poor combined radiation from all stars in the host
elliptical.  There is little differential rotation, and the lack of
shear may locally favour the formation of massive groups of luminous
stars in the disk affecting their surrounding ISM. On much smaller
sub-kiloparsec scales, the dynamics of a denser circumnuclear disk may
be only loosely related to those of the larger kiloparsec-sized
extended disk. Instead, in such a circumnuclear disk both dynamics and
excitation expected to be tied more closely to the properties of the
super-massive black hole in the nucleus.

\nobreak The emission lines from carbon and carbon monoxide are key to
understanding the properties of the ISM in galaxies, as they provide
almost all the cooling of the dense neutral gas.  Warm and tenuous gas
is traced by ionised carbon (\cii\,), warm and dense gas by neutral
carbon (\ci\,) and carbon monoxide (CO), and cold and dense gas by
(CO) and its isotopologues. Although the two neutral carbon \ci\ lines
at 492 and 809 GHz and CO lines with rest frequencies up to about 1000
GHz can be measured from the ground (albeit with increasing
difficulty), the ionised carbon \cii\ fine-structure line at 1.9 THz
requires a platform above most or all of the atmosphere.

\nobreak
NGC~5128 is the nearest ($D$ = 3.84 Mpc) giant elliptical (review by
Israel 1998). Its central black hole is still accreting in the
aftermath of a merger with a medium-sized late-type galaxy a few
hundred million years ago (Graham, 1979; Struve \etal 2010). The
remnant ISM of the merged galaxy has redistributed itself into a
warped, thin disk ('the extended thin disk' - hereafter called the ETD
- which shows in projection as a dark band crossing the optical image
of the galaxy (Dufour et al. 1979, Nicholson \etal 1992). It has
comparable amounts of atomic (\hi) and molecular H$_{2}$ gas, and its
total mass is uncertainly estimated at $1.5\times10^{9}$ M$_{\odot}$,
about two per cent of the enclosed dynamical mass.  On significantly
smaller scales, the nuclear black hole of $5\times10^{7}$ M$_{\odot}$
(Cappellari \etal 2009) is obscured by a compact circumnuclear disk
(CND) that feeds the black hole (Israel \etal 1990).

\nobreak Unlike the ETD, the compact CND is oriented at right angles
to the luminous radio and X-ray jets emanating from the Centaurus A
nucleus. This suggests that the CND and the jets are somehow
connected, but it is not yet clear how that may be.  The CND is a
well-defined entity with a diameter of $20''$ (400 pc)).  It has been
seen in FIR continuum (Hawarden \etal 1993) and $^{12}$CO line
emission (Espada \etal 2009), but was not yet studied in detail.

In this paper, we combine single-dish CO and [CI] observations of the
Cen A nucleus obtained from the ground over many years, and from
space. Although observations of the Cen~A nucleus distinguishing CND
and ETD in the lower ($J_{upper}\,\leq\,3$) transitions of CO are
found in a few earlier papers (Israel \etal 1990; Israel 1992;
Rydbeck \etal 1993), all data in this paper have much higher
signal-to-noise ratios and are presented here for the first time.

\section{Observations and data handling}

\subsection{Herschel-SPIRE}

\nobreak The nucleus of Centaurus A was observed with the Spectral and
Photometric Imaging Receiver and Fourier-Transform Spectrometer ({\it
  SPIRE-FTS} - Griffin \etal 2010) onboard the {\it Herschel} Space
Observatory\footnote{{\it Herschel} is an ESA space observatory with
  science instruments provided by European-led Principal Investigator
  consortia and with important participation from NASA} (Pilbratt et
al. 2010) in the single-pointing mode with sparse image sampling. The
FTS has two detector arrays (the SLW: wavelength range 303-671 $\mu$m
corresponding to a frequency range 447-989 GHz, and the SWW:
wavelength range 194-313 $\mu$m) corresponding to a frequency range
959-1544 GHz). The observations are summarised in
Table\,\ref{herschellog}.  The data were processed and calibrated
using {\it HIPE} version 9.0.0. The spectral range covered the $\co$
lines in the $J$=4-3 to $J$=13-12 transitions as well as the two
submillimeter \ci\ lines, which were all detected; however, no
$\thirco$ lines were detected.  The spectral resolution of 1.21 GHz
was insufficient to resolve any of the lines in the SLW spectrum, but
may start to resolve some of the lines in the SSW spectrum,
particularly the \nii\ line. Line fluxes were extracted by fitting a
sinc-gaussian function to the line profile. The beam FWHM values are
given in the on-line {\it Herschel-SPIRE} manual; they range from
29$''$ to 42$``$ for the SLW, and from 16.8$''$ to 21.1$''$ for the
SSW. At the overlap between the two arrays, at about 310 $\mu$m or 964
GHz, the beam size jumps from 21.1$''$ to 37.3$''$ going from SSW to
SLW.  The results are shown in Fig.\,\ref{spirespec}, and summarised
in Tables\,\ref{data12co} and \ref{datacarbon}.

\begin{table*}
\scriptsize
\caption[]{$\co$ observations}
\begin{center}
\begin{tabular}{lcccccccc}
\noalign{\smallskip}     
\hline
\noalign{\smallskip}
Transition& Frequency& Telescope & Beam   & Offset & Peak    & \multicolumn{2}{c}{$\int T_{mb} dV$} & $\int S_{\nu} dV$ \\
          &          &           & Size   &        & T$_{mb}$ & Integrated$^{a}$ &  Gauss Fit$^{b}$ & Integrated \\
          & (GHz)    &           & ($''$) & ($''$) & (mK)    & \multicolumn{2}{c}{($\kkms$)}       & (Jy $\kms$)\\
\noalign{\smallskip}     
\hline
\noalign{\smallskip}     
$J$=1-0   & 115.271  & {\it SEST} & 45 &    0 , 0     & 357 & 58   & 76   & 1096$\pm$164 \\
          &          &      &    &--16.4, +11.5 & 330 & 90   & 89   & 1701$\pm$170 \\
          &          &      &    &--32.8, +23.0 & 440 & 86   & 86   & 1625$\pm$163 \\
          &          &      &    & +16.4, --11.5& 450 & 88   & 84   & 1663$\pm$166 \\
          &          &      &    & +32.8, --23.0& 440 & 86   & 89   & 1625$\pm$170 \\
$J$=2-1   & 230.538  & {\it SEST} & 23 &    0 , 0     & 520 & 84   &104   & 1722$\pm$258 \\
          &          &      &    &--16.4, +11.5 & 540 & 62   & 63   & 1271$\pm$170 \\
          &          &      &    &--32.8, +23.0 & 500 & 61   & 61   & 1251$\pm$125 \\
          &          &      &    & +16.4, --11.5& 505 & 65   & 62   & 1333$\pm$133 \\
          &          &      &    & +32.8, --23.0& 440 & 62   & 59   & 1271$\pm$127 \\
$J$=3-2   & 345.796  & {\it JCMT} & 14 &    0 , 0     & 492 & 80   & 92   & 1481$\pm$222 \\
          &          &      &    &--18.8, +9.4  & 315 & 33   & --   &  602$\pm$90  \\
          &          &      &    &--34.5, +25.0 & 765 & 48   & --   &  875$\pm$131 \\
          &          &      &    &--40.7, +28.1 & 635 & 55   & --   & 1057$\pm$157 \\
          &          &      &    & +18.4, --9.4 & 875 & 60   & --   & 1094$\pm$164 \\
          &          &      &    & +34.5, --25.0& 635 & 61   & --   & 1167$\pm$175 \\
          &          &      &    & +40.7, --28.1& 365 & 30   & --   &  547$\pm$82  \\
$J$=4-3   & 461.041  & {\it APEX} & 14 &    0 , 0     & 287 & 57   & 64   & 1651$\pm$330 \\
          &          & {\it SPIRE}& 41 &    0 , 0     &     &      &      & 7364$\pm$638 \\
$J$=5-4   & 576.268  & {\it SPIRE}& 33 &    0 , 0     &     &      &      & 5196$\pm$234 \\
          &          & {\it HIFI} & 38 &    0 , 0     &  75 & 13.7 & 14.6 & 4709$\pm$471 \\
          &          &      &    &--7.5 , +7.5  &  65:& 12   & ---  &  --- \\
          &          &      &    & +7.5 , --7.5 &  73:& 13   & ---  &  --- \\   
$J$=6-5   & 691.473  & {\it APEX} &  9 &    0 , 0     &  99 & 27   & 32   &  787$\pm$187 \\
          &          & {\it SPIRE}& 29 &    0 , 0     &     &      &      & 2768$\pm$88  \\
          &          & {\it HIFI} & 33 &    0 , 0     &  30 &  9.9 & 10.3 & 3460$\pm$573 \\
          &          &      &    &--7.5 , +7.5  &  48:&  8:  & ---  & \\
          &          &      &    & +7.5 , --7.5 &  48:&  8:  & ---  & \\   
$J$=7-6   & 806.652  & {\it SPIRE}& 35 &    0 , 0     &     &      &      & 1977$\pm$70  \\
          &          & {\it HIFI} & 27 &    0 , 0     &  20 &  4.4 &  4.7 & 1548$\pm$232 \\
          &          &      &    &--7.5 , +7.5  &  35:&  2:  & ---  &  --- \\
          &          &      &    & +7.5 , --7.5 &  43:&  5:  & ---  &  --- \\   
$J$=8-7   & 921.799  & {\it SPIRE}& 36 &    0 , 0     &     &      &      & 1548$\pm$102 \\
          &          & {\it HIFI} & 25 &    0 , 0     &  21 &  4.7 &  5.2 & 1669$\pm$419 \\
          &          &      &    &--7.5 , +7.5  &  32:&  3:  & ---  &  --- \\
          &          &      &    & +7.5 , --7.5 &  31:&  2:  & ---  &  --- \\   
$J$=9-8   & 1036.912 & {\it SPIRE}&18.7&    0 , 0     &     &      &      &  632$\pm$71  \\ 
          &          & {\it HIFI} & 23 &    0 , 0     &  14 &  2.7 &  2.9 & 1259$\pm$212 \\ 
          &          &      &    & -7.5 , +7.5  &  50:&  4:  & ---  &  --- \\
          &          &      &    & +7.5 , --7.5 &  34:&  7:  & ---  &  --- \\   
$J$=10-9  & 1151.986 & {\it SPIRE}&17.1&    0 , 0     &     &      &      &  341$\pm$62  \\ 
          &          & {\it HIFI} & 20 & average 3 pos &$<14$&  0.5:& ---  &  281$\pm$93  \\ 
$J$=11-10 & 1267.014 & {\it SPIRE}&17.6&    0 , 0     &     &      &      &  306$\pm$68  \\ 
$J$=12-11 & 1381.995 & {\it SPIRE}&16.9&    0 , 0     &     &      &      &  183$\pm$57  \\ 
$J$=13-12 & 1496.923 & {\it SPIRE}&16.8&    0 , 0     &     &      &      &  393$\pm$64  \\ 
          &          & {\it HIFI} & 15 & average 3 pos &$<40$& $<1.2$& ---& $<435$ \\ 
\noalign{\smallskip}     
\hline
\noalign{\smallskip}
\end{tabular}
\end{center}
Notes: (a) Summation over all amplitudes in the velocity interval 
300-800 km s$^{-1}$, i.e. the integrated emission line intensity was 
not corrected for absorption.  (b) Sum of one or more gaussians 
fitted to the observed line profile, excluding the velocity range 
between 500 $\kms$ and 625 $\kms$, i.e the integrated emission line 
intensity is to first order unaffected by absorption.
\label{data12co}
\end{table*}

\begin{table*}
\scriptsize
\caption[]{$\thirco$ and HCN observations}
\begin{center}
\begin{tabular}{lccccccccccc}
\noalign{\smallskip}     
\hline
\noalign{\smallskip}
Transition&Frequency& Telescope & Beam   & Offset  & Peak    & $\int T_{mb} dV$ & $\int S_{\nu} dV$ & \multicolumn{4}{c}{Ratio $\co$/$\thirco$}$^{a}$ \\
          &         &           & Size   &         & T$_{mb}$ &                 & Integrated                 & All   & ETD   &  CND SE & CND NW \\
          & (GHz)   &           & ($''$) & ($''$)  & (mK)    &($\kkms$)        & (Jy $\kms$ )     &       &        &         &        \\
\noalign{\smallskip}     
\hline
\noalign{\smallskip}     
\multicolumn{12}{c}{\bf $\thirco$}\\
\noalign{\smallskip}     
\hline
\noalign{\smallskip}
$J$=1-0   & 110.201 & {\it SEST} & 47 &    0 , 0     & 39 & 7.4 & 140$\pm$28 &  9$\pm$2 & 10$\pm$2 & 10$\pm$4 & 11$\pm$4 \\
          &         &      &    & --16, +12    & 34 & 7.5 & 142$\pm15$ &          &          &            &  \\
          &         &      &    & --32, +24    & 23 & 3.7 &  70$\pm$11 & 23       &          &            &  \\
          &         &      &    & --40, +28    & 19 & 2.7 &  51$\pm$8  & --       &          &            &  \\
          &         &      &    & --48, +32    & 13 & 1.8 &  34$\pm$5  & --       &          &            &  \\
          &         &      &    &  +16, --12   & 36 & 6.5 & 123$\pm$18 &          &          &            &  \\
          &         &      &    &  +32, --24   & 26 & 3.8 &  72$\pm$11 & 23       &          &            &  \\
          &         &      &    &  +40, --32   & 19 & 3.1 &  59$\pm$9  & --       &          &            &  \\
          &         &      &    &  +48, 032    & 18 & 2.1 &  40$\pm$6  & --       &          &            &  \\
$J$=2-1   & 220.399 & {\it SEST} & 24 &    0 , 0     & 46 & 7.8 & 160$\pm$32 & 11$\pm$1 & 13$\pm$2 & 13$\pm$1.5 & 11$\pm$1.5 \\
          &         &      &    &--16.4, +11.5 & 38 & 5.4 & 111$\pm$17 & 12       &          &            &  \\
          &         &      &    &--32.8, +23.0 & 54 & 7.4 & 152$\pm$23 &  8       &          &            &  \\
          &         &      &    &--41.0, +28.5 & 36 & 4.2 &  86$\pm$13 & --       &          &            &  \\
          &         &      &    & +16.4, --11.5& 52 & 8.2 & 168$\pm$25 & 7.5      &          &            &  \\
          &         &      &    & +32.8, --23.0& 50 & 7.1 & 146$\pm$22 & 8.5      &          &            &  \\
          &         &      &    & +41.0, --28.5& 48 & 7.0 & 144$\pm$22 & --       &          &            &  \\
          &         &      &    & +49.2, --34.2& 48 & 6.6 & 135$\pm$20 & --       &          &            &  \\
$J$=3-2   & 330.588 & {\it JCMT} & 15 &    0 , 0     & 41 & 6.2 & 182$\pm$36 & 14$\pm2$ & 14$\pm$2 & 17$\pm$4   & 13$\pm$2 \\ 
          &         & {\it APEX} & 18 &    0 , 0     & 87 & 14  & 419$\pm$84 &          &          &            &  \\ 
          &         &      &    &--20.8, +12.0 & 41 & 6.3 & 189$\pm$28 &          &          &            &  \\
          &         &      &    & +20.8, --12.0& 24 & 4.4 & 132$\pm$20 &          &          &            &  \\
$J$=5-4   & 550.926 &{\it  HIFI} & 45 &   average   & 1.2 & 0.7 & 244$\pm$85 & 18$\pm$6 & ----     & ----       & ---- \\
\noalign{\smallskip}     
\hline
\noalign{\smallskip}
\multicolumn{12}{c}{\bf HCN}\\
\noalign{\smallskip}     
\hline
\noalign{\smallskip}
$J$=1-0   &  88.632 &{\it SEST} & 57 &     0 , 0    & 0.026  & 1.60$^{b}$ & --- & --- \\
          &         &           &    & -16.4, +11.4 & 0.015  & 0.59$^{b}$ & --- & --- \\
          &         &           &    & -49.2, +34.2 &$<$0.008& 0.24       & --- & --- \\
          &         &           &    & +16.4, -11.4 & 0.015  & 0.20$^{b}$ & --- & --- \\
          &         &           &    & +49.2, +34.2 &$<$0.008& 0.25       & --- & --- \\
\noalign{\smallskip}     
\hline
\noalign{\smallskip}
\end{tabular}
\end{center}
Notes: (a) Ratios determined by (partial) fitting $\thirco$ profile
to $\co$ profile; (b) Profiles dominated by absorption; integrated values cover velocity ranges $V_{LSR}$ = 300-500 $\kms$ and $V_{LSR}$ = 600-800 $\kms$ only.
\label{data13co}
\end{table*}

\begin{table*}
\scriptsize
\caption[]{C$^0$ and C$^+$ Observations}
\begin{center}
\begin{tabular}{lcccccccc}
\noalign{\smallskip}     
\hline
\noalign{\smallskip}
Transition & Frequency & Telescope & Beam   & Offset & Peak     & \multicolumn{2}{c}{$\int T_{mb}$ dV} & $\int S_{\nu} dV$ \\
           &           &           & Size   &         & T$_{mb}$ & Integrated$^{a}$ & Gauss Fit$^{b}$ & Integrated \\
           & (GHz)     &           & ($''$) & ($''$)  & (K)      &\multicolumn{2}{c}{($\kkms$)}        & (Jy $\kms$)      \\
\noalign{\smallskip}     
\hline
\noalign{\smallskip}     
\ci $J$=1-0 & 492.161 &{\it  APEX} & 12.5 &   0 , 0    & 342 &  84.6 & 100 &  2880$\pm$432 \\ 
            &         &{\it SPIRE} & 38   &   0 , 0    &     &       &     &  7014$\pm$328 \\
            &         & {\it HIFI} & 44.1 &   0 , 0    &  79 &  17.0 &  19 &  6700$\pm$1005\\
            &         &      &      &--7.5, +7.5 &  86 &  18.3 &  20 &  7053$\pm$1058\\
            &         &      &      & +7.5, --7.5&  78 &  14.8 &  16 &  5642$\pm$846 \\
\ci $J$=2-1 & 809.342 & {\it APEX} &  7.7 &   0 , 0    & 510 & 104.3 & 136 &  4094$\pm$614 \\ 
            &         &{\it SPIRE} & 34   &   0 , 0    &     &       &     & 11196$\pm$80  \\
            &         & {\it HIFI} & 26.5 &   0 , 0    &  99 &  20.7 &  26 &  9146$\pm$1372\\
            &         &      &      &--7.5, +7.5 &  81 &  18.4 &  21 &  7387$\pm$1108\\ 
            &         &      &      & +7.5, --7.5&  91 & 14.7  &  18 &  6332$\pm$950 \\
\nii\       & 1426.1  &{\it SPIRE} & 16.9 &   0 , 0    &     &       &     & 10090$\pm$97  \\
\cii        & 1900.539& {\it HIFI} & 11.1 &   0 , 0    & 885 & 140.0 & 144 & 50276$\pm$7541\\
            &         &      &      &--7.5, +7.5 & 790 &  60.8 &  59 & 21228$\pm$3184\\  
            &         &      &      & +7.5, --7.5&1290 & 102.3 &  95 & 33168$\pm$4975\\
\noalign{\smallskip}     
\hline
\noalign{\smallskip}
\end{tabular}
\end{center}
Notes: (a) Summation over all amplitudes in the velocity interval 
300-800 km s$^{-1}$, i.e. the integrated emission line intensity was 
not corrected for absorption. In the case of the [CII] profiles, the 
velocity interval was 370 to 720 km s$^{-1}$. (b) Sum of one or
more gaussians fitted to the observed line profile, excluding the
velocity range between 500 $\kms$ and 625 $\kms$, i.e the integrated
emission line intensity is to first order unaffected by absorption.
\label{datacarbon}
\end{table*}

\begin{figure*}[]
\centering
\includegraphics[width=18cm,clip]{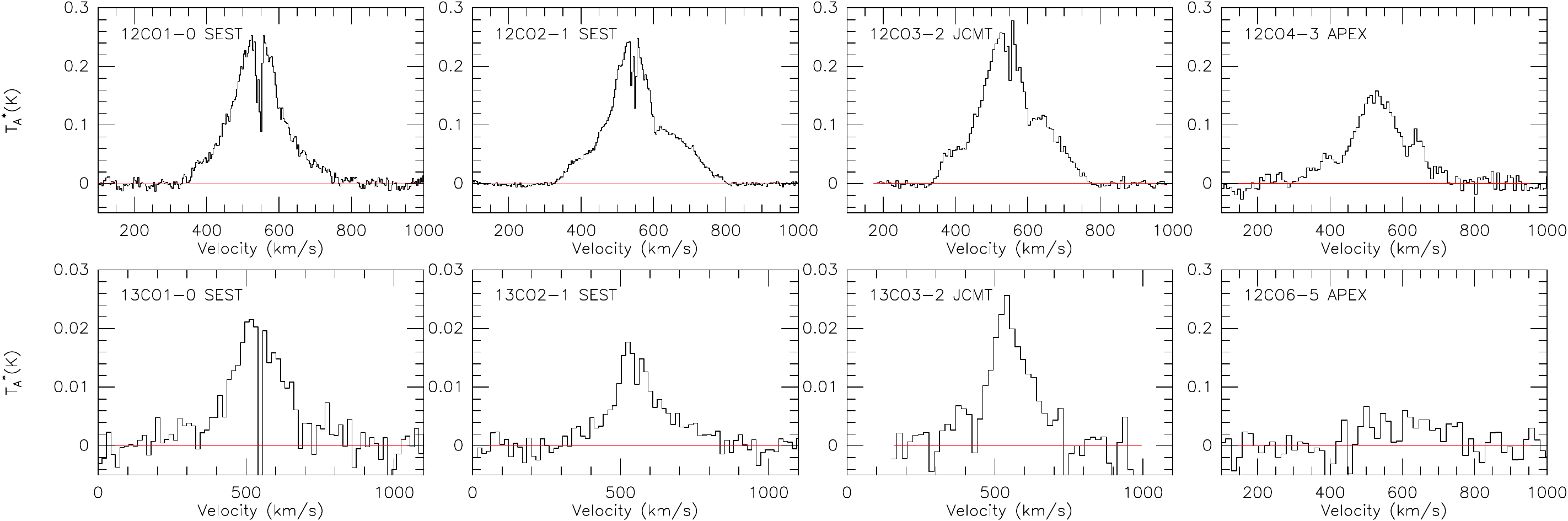}
\includegraphics[width=18cm,clip]{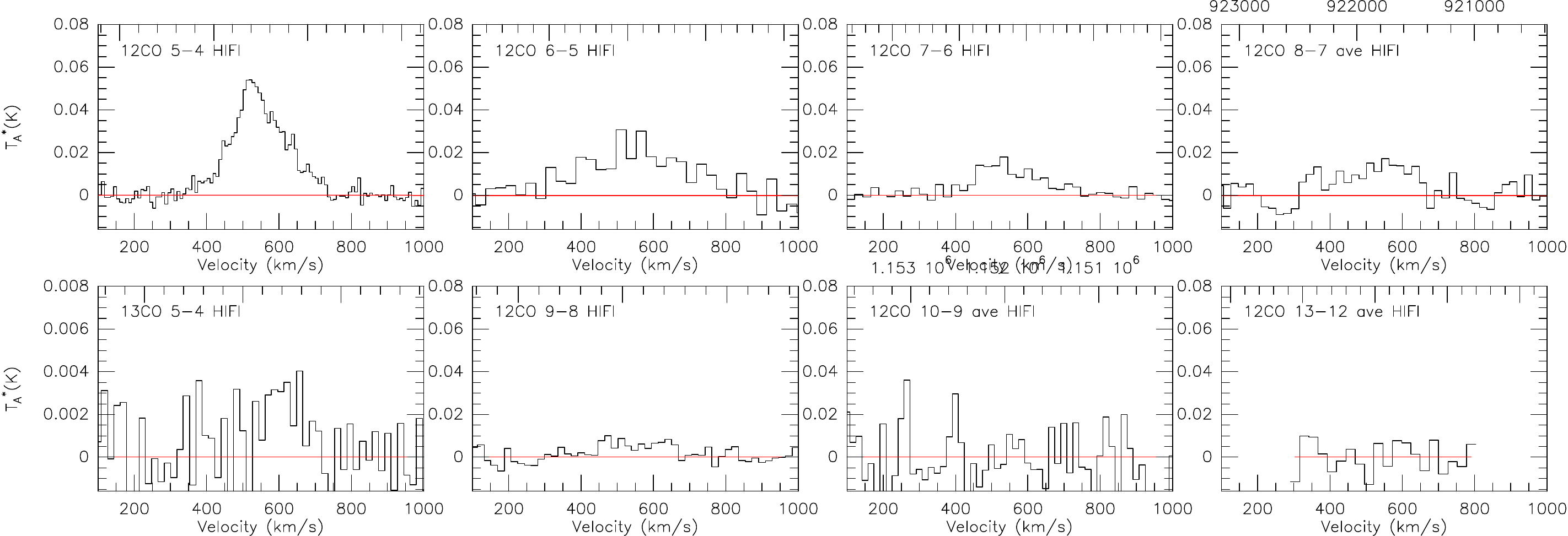}
\caption[]{Baseline-subtracted line profiles towards the center of
  NGC~5128 (Centaurus A) obtained with the ground-based telescopes
  {\it SEST}, {\it JCMT}, and {\it APEX} (top two rows) and with the
  HIFI instrument on-board of the Herschel Space Observatory (bottom
  two rows).  Vertical scale is T$_{\rm A}^{*}$ in Kelvin, horizontal
  scale is V$_{\rm LSR}$ in $\kms$. Species, transition and telescope
  used are identified at the top left corner of each panel. The HIFI
  $\co$ results for the $J$=8-7, $J$=10-9, and $J$=13-12 transitions
  are the averages of the profile towards the nucleus and the adjacent
  positions separated from that by $10''$. Note that the deep
  absorption apparent in the $^{13}$CO (1-0) profile is real; its
  depth to negative values reflects the weakness of the line emission
  relative to the (subtracted) continuum strength at the same
  frequency. For more details, see Section 2 and
  Tables\,\ref{data12co} and \ref{data13co}.  }
\label{cofig}
\end{figure*}

\begin{figure*}[]
\centering
\includegraphics[width=18cm, clip]{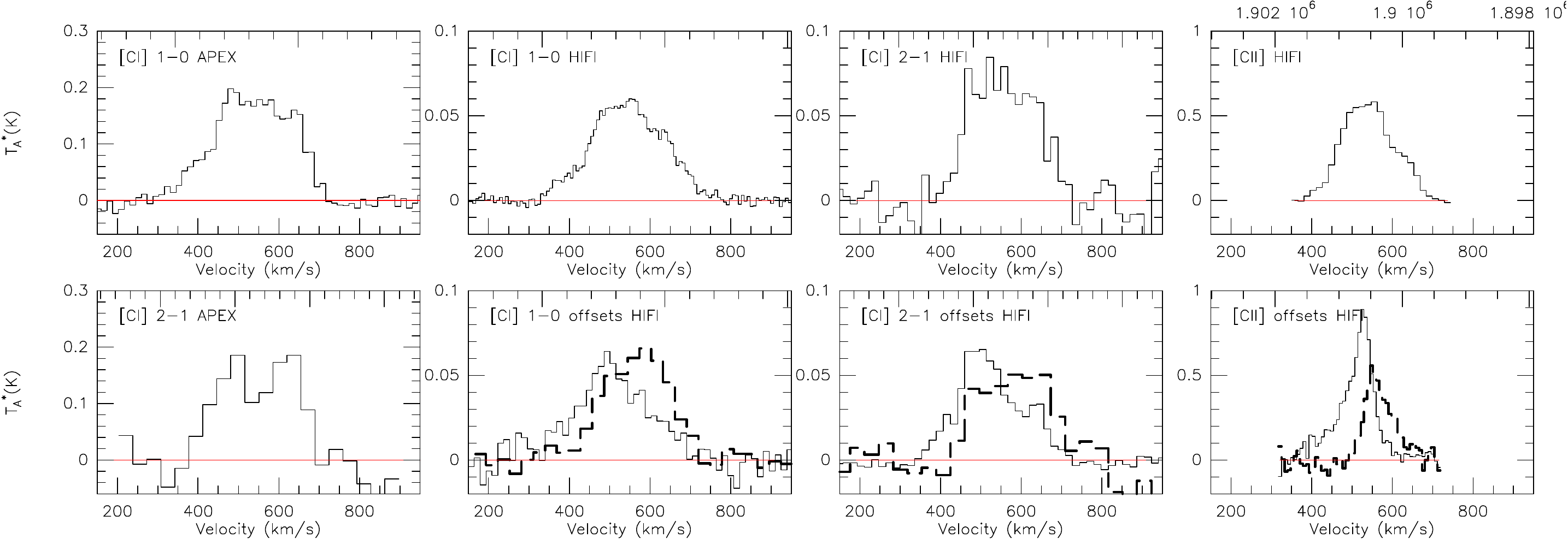}
\caption[]{Baseline-subtracted line profiles towards the center of
  NGC~5128 (Centaurus A) obtained with the APEX telescope and the HIFI
  instrument on-board the Herschel Space Observatory.  The HIFI \ci
  and \cii profiles for the nuclear position, and the positions offset
  by $\pm\,10''$ from the nucleus, are shown separately. The NW offset
  profile is represented by a dashed line, the SE offset profile by a
  solid line. Species, transition and telescope used are identified at
  the top left corner of each panel. Vertical scale is T$_{\rm A}^{*}$
  in Kelvin, horizontal scale is V$_{\rm LSR}$ in $\kms$. For more
  details, see Table\,\ref{datacarbon}.  }
\label{carbonfig}
\end{figure*}

\subsection{Herschel-HIFI}

\nobreak The center of Centaurus A was also observed with the
Heterodyne Instrument for the Far Infrared ({\it HIFI} - de Graauw et
al. 2010) on-board {\it Herschel} as part of the Guaranteed Time Key
Programme {\it HEXGAL} (PI: R. G\"usten). Observations were carried
out in fast-chopping dual-beam switch mode using a wobbler throw of
3$'$ for all observations, and are summarised in
Table\,\ref{herschellog}.  Calibration was achieved through hot-cold
absorber measurements.  The data were recorded using the wide-band
acousto-optical spectrometer, consisting of four units with a
bandwidth of 1 GHz each, covering the 4 GHz IF for each polarisation
with spectral resolutions of 1 MHz. Data were reduced using the {\it
  HIPE} and {\it CLASS} software packages.  For each scan, we combined
the four sub-bands in each polarisation to create a 4 GHz spectrum. We
subtracted first-order baselines. For each line, we inspected the
result in each polarisation (H and V) separately.  The continuum and
line amplitudes agreed within 15$\%$. We have used the calibration in
Table 5.5 of the on-line {\it HIFI} Observer's Manual to convert
antenna temperatures to main-beam temperatures and flux
densities. Between 480 and 960 GHz, the main-beam efficiency is almost
constant, dropping from $\eta_{\rm mb}\,=\,0.76$ to $\eta_{\rm
  mb}\,=\,0.74$, and the antenna temperature to flux conversion factor
likewise changes only little from 464 to 472 Jy/K. 

\begin{figure*}[]
\centering
\includegraphics[width=10cm,clip]{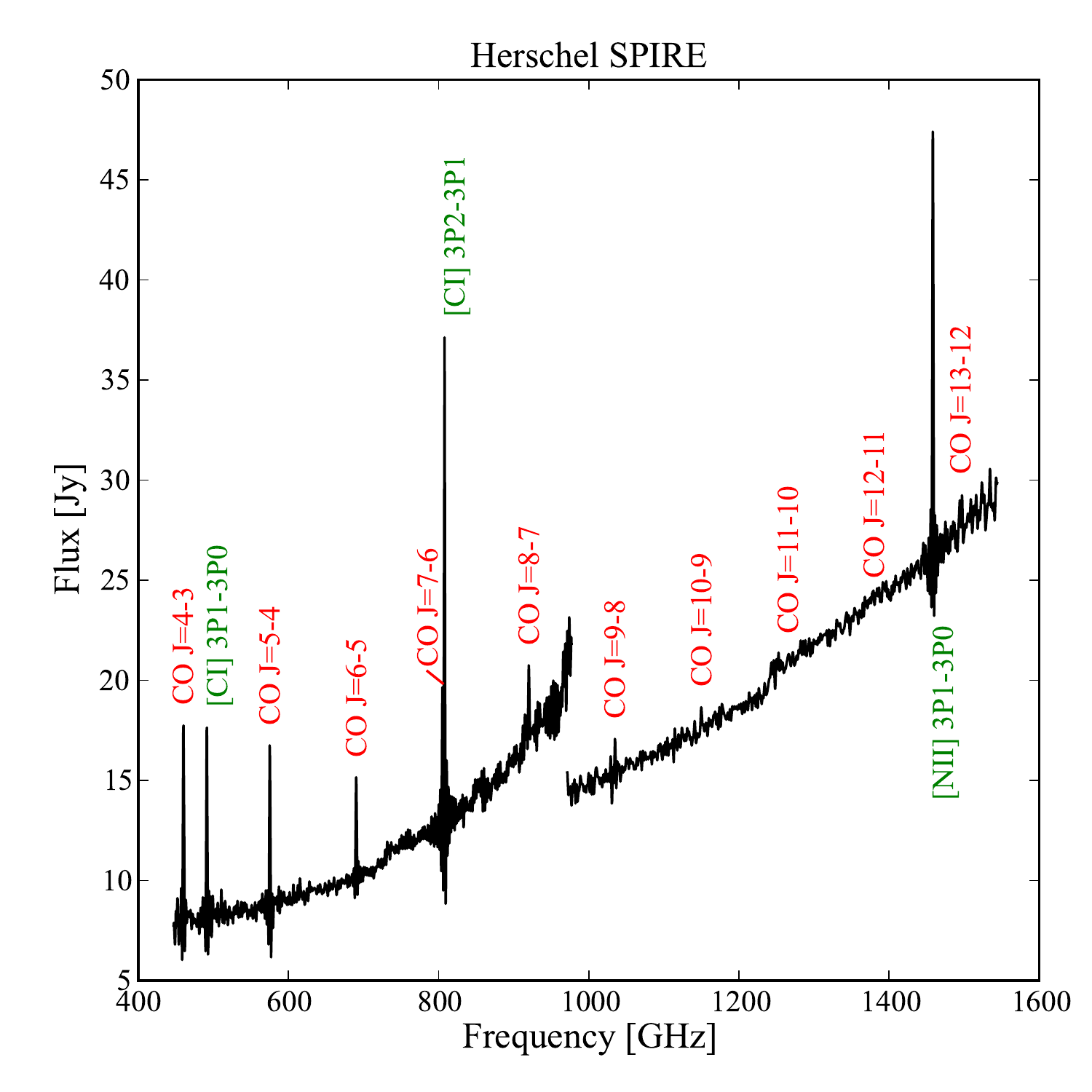}
\includegraphics[width=8cm,clip]{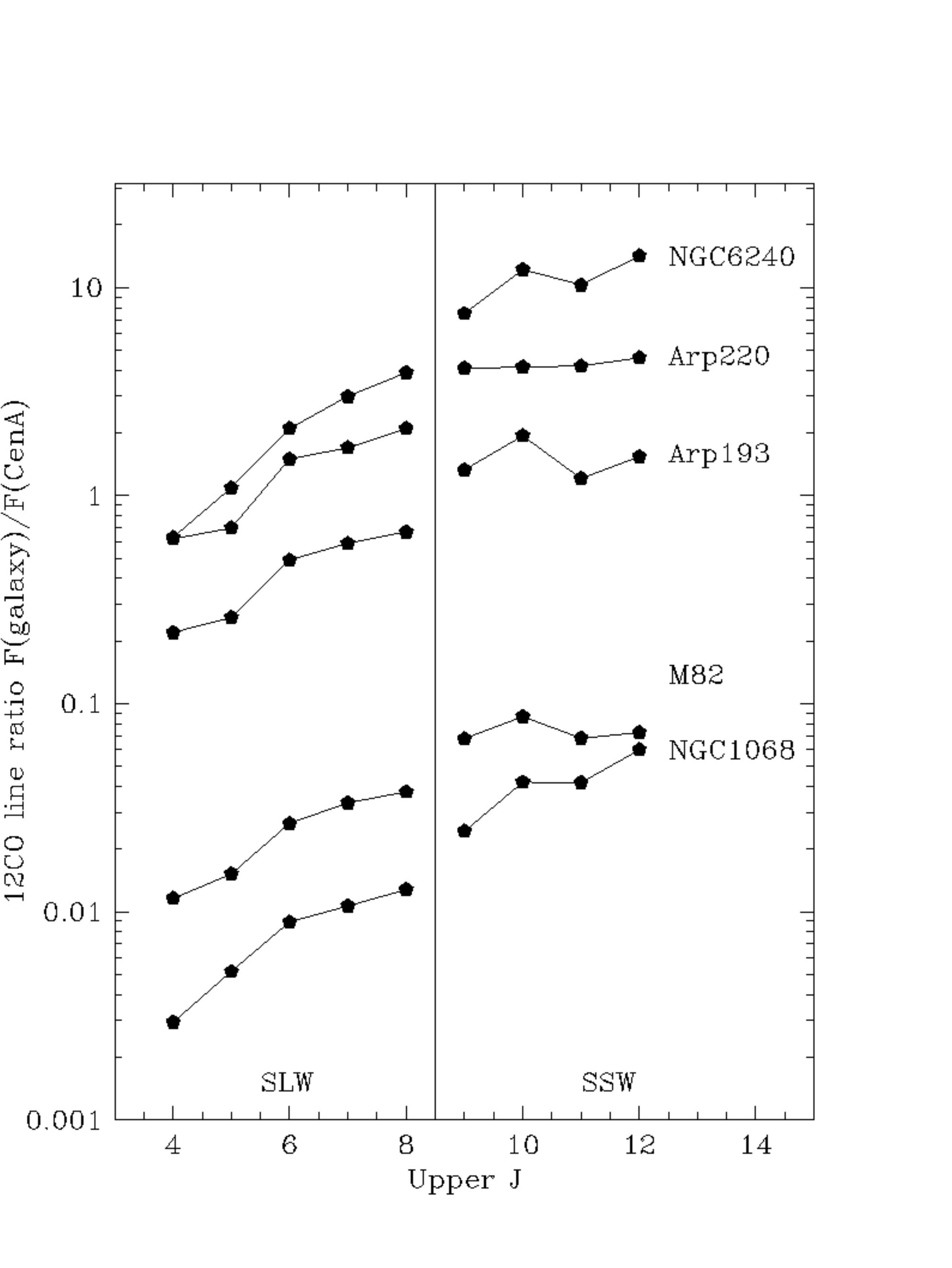}
\caption[]{Left: full submillimeter spectrum of the center of NGC~5128
  (Centaurus A) obtained with the {\it SPIRE} instrument on board the
  Herschel Space Observatory (See Sect. 2).  Species and transitions
  are identified throughout. The jump in the continuum at 944 GHz is
  caused by the different angular resolutions of the SLW and the SSW
  (see Section 2).  The {\it SPIRE} continuum contains a contribution
  from the point source nucleus of about 8.2 Jy at 460 GHz, slowly
  decreasing with frequency. The remaining continuum is due to
  extended thermal emission from dust, increasing with frequency.
  Vertical scale is flux in Jansky, horizontal scale is frequency in
  GHz. For more details, see Tables\,\ref{data12co} and
  \ref{data13co}.  Right: comparison of {\it SPIRE} CO line fluxes of
  the NGC~5128 center with those of the starburst galaxy M~82, the
  AGN+starburst NGC~1068, and the (U)LIRGs Arp 193, Arp 220, and
  NGC~6240. Mrk~231 has not been marked separately as its spectral
  ladder is identical to that of NGC~6240. The ratio of the galaxy CO
  line flux to the Cen~A CO line flux is shown for each $J$
  transition, as observed by {\it SPIRE} without correction for finite
  beam-size, source extent, and beam efficiency. The vertical line
  separates measurements obtained with the SSW from those obtained
  with the SLW in a beam roughly twice as wide.}
\label{spirespec}
\end{figure*}

\begin{table}
\scriptsize
\caption[]{Normalized spectral line emission distribution}
\begin{center}
\begin{tabular}{lccccc}
\noalign{\smallskip}     
\hline
\noalign{\smallskip}
Transition & Obs.  & $BNF^{a}$ & $F_{em}^{b}$ & $F_{abs}^{c}$ & $L_{cor}^{d}$\\
           &       &          & \multicolumn{2}{c}{(Jy $\kms$)} & ($10^{4}$ L$_{\odot}$)\\
\noalign{\smallskip}     
\hline
\noalign{\smallskip}     
\multicolumn{6}{c}{The $\co$ ladder} \\
\noalign{\smallskip}     
\hline
\noalign{\smallskip}     
 $J$=1-0   & {\it SEST}  &   0.53   &  583$\pm$87  & 340 & 0.16$\pm$0.05 \\ 
 $J$=2-1   & {\it SEST}  &   0.99   & 1705$\pm$256 & 410 & 0.72$\pm$0.20 \\ 
 $J$=3-2   & {\it JCMT}  &   2.13   & 3151$\pm$473 & 220 & 1.71$\pm$0.29 \\ 
 $J$=4-3   & {\it APEX}  &   2.13   & 3513$\pm$703 & 195 & 2.83$\pm$0.61 \\ 
           & {\it SPIRE} &   0.57   & 4208$\pm$365 &     & \\ 
 $J$=5-4   & {\it SPIRE} &   0.66   & 3441$\pm$155 & 170 & 2.87$\pm$0.20 \\ 
           & {\it HIFI}  &   0.60   & 2803$\pm$280 &     & \\ 
 $J$=6-5   & {\it SPIRE} &   0.75   & 2066$\pm$66  & 145 & 2.30$\pm$0.15 \\ 
           & {\it HIFI}  &   0.66   & 2291$\pm$380 &     & \\ 
 $J$=7-6   & {\it SPIRE} &   0.63   & 1267$\pm$46  & 140 & 1.58$\pm$0.13 \\ 
           & {\it HIFI}  &   0.79   & 1229$\pm$185 &     & \\ 
 $J$=8-7   & {\it SPIRE} &   0.62   &  961$\pm$63  &     & 1.48$\pm$0.20 \\ 
           & {\it HIFI}  &   0.89   & 1477$\pm$370 &     & \\ 
 $J$=9-8   & {\it SPIRE} &   1.25   &  790$\pm$89  &     & 1.44$\pm$0.16 \\ 
           & {\it HIFI}  &   0.99   & 1246$\pm$210 &     & \\ 
 $J$=10-9  & {\it SPIRE} &   1.45   &  494$\pm$90  &     & 0.76$\pm$0.15 \\ 
           & {\it HIFI}  &   1.14   &  319$\pm$106 &     & \\ 
 $J$=11-10 & {\it SPIRE} &   1.39   &  425$\pm$120 &     & 0.79$\pm$0.22 \\ 
 $J$=12-11 & {\it SPIRE} &   1.47   &  269$\pm$90  &     & 0.55$\pm$0.18 \\ 
\noalign{\smallskip}     
\hline
\noalign{\smallskip}     
\multicolumn{6}{c}{The $\thirco$ ladder} \\
\noalign{\smallskip}     
\hline
\noalign{\smallskip}     
 $J$=1-0   &  &  &  &  & 0.017$\pm$0.007 \\ 
 $J$=2-1   &  &  &  &  & 0.065$\pm$0.019 \\ 
 $J$=3-2   &  &  &  &  & 0.122$\pm$0.027 \\ 
 $J$=4-3   &  &  &  &  &      ---        \\ 
 $J$=5-4   &  &  &  &  & 0.16$\pm$0.0.04 \\ 
\noalign{\smallskip}     
\hline
\noalign{\smallskip}     
\multicolumn{6}{c}{The carbon $\c$ and $\cp$ lines} \\
\noalign{\smallskip}     
\hline
\noalign{\smallskip}     
\ci (1-0)  & {\it APEX}  & 2.17     & 6149$\pm$922 & 655 &   4.2$\pm$0.8 \\
           & {\it SPIRE} & 0.63     & 4418$\pm$207 &     &               \\
           & {\it HIFI}  & 0.58     & 5360$\pm$804 &     &               \\
\ci (2-1)  & {\it APEX}  & 7:       &     ---      &1782 &  11.1$\pm$2.4 \\
           & {\it SPIRE} & 0.68     & 7613$\pm$761 &     &               \\
           & {\it HIFI}  & 0.81     & 7408$\pm$1111&     &               \\
\cii\      & {\it HIFI}  & 2.8     &140773$\pm$21115&1436& 397$\pm$60    \\ 
\nii\      & {\it SPIRE} & 1.47     &14832$\pm$2000& --- &  31$\pm$5     \\
\noalign{\smallskip}     
\hline
\noalign{\smallskip}
\end{tabular}
\end{center} 
Notes: (a) Beam normalisation factor (BNF) to reduce observed emission
line fluxes from Table\,ref{data12co}, \ref{data13co}, \ref{datacarbon}
to those enclosed by a $22''$ beam; for line fluxes expressed in $\kkms$, 
multiply BNF by $(\theta_{b}/22)^{2}$, where $\theta_{b}$ is the observing 
FWHM beam-size from Tables\,\ref{data12co} and \ref{datacarbon}.  
(b) Emission line fluxes normalised to the response of a $22''$
beam by application of the BNF. 
(c) Beam-independent absorption line flux derived from the
estimated difference between integrated and gaussian-fitted line
intensities in Table\,\ref{data12co}; adopted error: $50\%$. Note 
that this is the total amount of absorption over the whole profile, 
not just the prominently visible deep absorption at systemic 
velocities. 
(d) Final adopted emission line luminosity in a $22''$ beam, corrected 
for absorption line losses. For $\thirco$ we did not determine line fluxes, 
but derived corrected luminosities directly from the more accurate $\co$ 
values using the mean of the $\co/\thirco$ ratios in 
Table\,\ref{data13co}.
\label{luminosities}
\end{table}

\nobreak With the exception of the C$^{18}$O(5-4) and \nii\ lines
(neither of which was detected), we observed all lines at three
positions: the center, and two positions offset by 10$''$ to either
side in a position angle of 45$^{\rm o}$ in order to separately sample
emission from the circumnuclear disk. The central position was
observed with reasonably good signal-to-noise ratios in the $J$=5-4,
$J$=7-6, $J$=9-8, and $J$=13-12 transitions of $\co$, as were the
$J$=13-12 $\co$ NW and SE offset positions. The other CO measurements
had much shorter durations resulting in poorly defined spectral
profiles, but the \ci\ and \cii\ lines were again observed with (very)
good signal-to-noise ratios at all positions.  As shown in
Tables\,\ref{data12co}, \ref{data13co}, and \ref{datacarbon}, the {\it
  HIFI} beam sizes ranged from $38\arcsec$ at 576 GHz to $11\arcsec$
at 1900 GHz. The profiles observed with {\it HIFI} are shown in
Fig.\,\ref{cofig} and \ref{carbonfig}.

\subsection{APEX 12m}

\nobreak Between 2007 and 2011 we have used the Vertex Antennentechnik
{\it APEX} \footnote{The Atacama Pathfinder Experiment (APEX) is a
  collaboration between the Max-Planck-Institut für Radioastronomie
  (MPIfR), the European Southern Observatory (ESO), and the Onsala
  Space Observatory (OSO).}  12-m telescope (G\"usten \etal 2006) to
observe the nucleus of NGC~5128 in the $J$=3-2 $\thirco$ transition at
330 GHz, the $J$=4-3 and $J$=6-5 $\co$ transitions at 461 and 691 GHz
respectively, and the two submillimeter \ci\ transitions at 492 and
809 GHz. The location of {\it APEX} at the high elevation of 5105 m
renders it very suitable to high-frequency observations from the
ground. The observations were made with the First Light {\it APEX}
Submillimeter Heterodyne ({\it FLASH}) dual-frequency receiver
(Heyminck \etal 2006) and the Carbon Heterodyne Array ({\it CHAMP+})
receiver (G\"usten \etal 2008; Kasemann \etal 2006), both developed by
the Max Planck Institut f\"ur Radioastronomie in Bonn
(Germany). Main-beam efficiencies were 0.73, 0.60, 0.56, and 0.43 at
operating frequencies of 352, 464, 650, and 812 GHz. At the same
frequencies, the antenna temperature to flux density conversion
factors were 41, 48, 53, and 70 Jy/K, respectively, with beam sizes
ranging from $18\arcsec$ to $7.7\arcsec$.

\nobreak All observations were done under excellent weather conditions
with typical overall system temperatures of 2100 K for {\it CHAMP+-I}
(SSB, 690 GHz), 7500 K for{\it CHAMP+-II} (SSB, 800 GHz), and 1100 K
for {\it FLASH-I} (DSB, 460 GHz), 290 K for {\it APEX-1} and 230 K for
{\it APEX-2a} (both SSB). Calibration errors are estimated at 15 to
20$\%$. Observations were taken using Fast Fourier Transform
Spectrometer (FFTS) (Klein et al. 2006) back-ends for all instruments,
except {\it CHAMP+}, for which only the two central pixels were
attached to the FFTS back-ends. Other {\it CHAMP+} pixels were attached
to the MPI Array Correlator System (MACS) back-ends. FFTS back-ends are
able to reach resolutions of 0.12 MHz (0.045 $\kms$ at 800 GHz), while
the MACS units were used at a resolution of 1 MHz (0.36 $\kms$ at 800
GHz). {\it APEX} absolute pointing accuracy is $\sim 2''$ (r.m.s.),
but its pointing on track is accurate to $0.6''$ (r.m.s.). All
observations were taken using position switching with reference
positions in azimuth ranging from 600$''$ to 3600$''$.  The results
are shown in Fig.\,\ref{cofig} and \ref{carbonfig}, and summarised in
Tables\,\ref{data12co},\,\ref{data13co}, and \ref{datacarbon}.

\subsection{JCMT 15m}

\nobreak The 15m James Clerk Maxwell Telescope ({\it
  JCMT})\footnote{The The James Clerk Maxwell Telescope is operated by
  the Joint Astronomy Centre on behalf of the Science and Technology
  Facilities Council of the United Kingdom, the National Research
  Council of Canada, and (until 31 March 2013) the Netherlands
  Organisation for Scientific Research.} on top of Mauna Kea (Hawaii)
was used between 2003 and 2005 to measure the $J$=3-2 transitions of
$\co$ and $\thirco$ at 345 and 330 GHz respectively towards the
nucleus of Centaurus A.  At the observing frequencies, the beam size
was 14$''$ and the main-beam efficiency was 0.62. The antenna
temperature to flux density conversion was 29.4 Jy/K. We used the
dual-polarisation receiver RxB and the Digital Autocorrelating
Spectrometer (DAS) back-end in wide-band mode, with total band-widths
of 920 MHz (800 km s$^{-1}$) and 500 MHz (435 $\kms$, providing
velocity resolutions of 0.65 and 0.325 $\kms$ respectively.  As seen
from Hawaii, Centaurus A culminates at an elevation of only 23$^{\rm
  o}$ above the horizon. We observed the object only when above
20$^{\rm o}$, i.e. about 2 hours per day. All observations were taken
in beam-switching mode with a throw of 3$'$ in azimuth.  The results
are shown in Fig.\,\ref{cofig}, and summarised in
Tables\,\ref{data12co} and \ref{data13co}.

\subsection{SEST 15m}

\nobreak We have used the 15m Swedish ESO Submillimetre Telescope
({\it SEST})\footnote{The Swedish-ESO Submillimeter Telescope was
  operated jointly by the European Southern Observatory (ESO) and the
  Swedish Science Research Council (NFR)} (Booth \etal 1989) on top of
Cerro La Silla (Chile) to observe the $J$=1-0 HCN, $J$=1-0, and
$J$=2-1 $\co$ and $\thirco$ transitions with angular resolutions of
57$"$, 45$''$, and 23$''$ respectively. In order to convert observed
antenna temperatures to main-beam temperatures, we use efficiencies
$\eta_{\rm mb}$ = 0.75, 0.70, and 0.50 respectively. Similarly, we
derive flux densities from the observed antenna temperatures by
applying the respective conversion factors 25, 27, and 41 Jy/K. All
observations were made in the double-beam switching mode, with a throw
of 12$'$ and a frequency of 6 Hz, producing excellent baseline
stability. In the early period 1989-1993, we used a relatively noisy
Schottky barrier diode receiver. Between 1996 and 2003 we obtained
spectra with the more sensitive SIS receiver, using high- and
low-resolution AOS back-ends in parallel. The data presented in this
paper were taken with the latter, which had a total bandwidth of 500
MHz (later 1 GHz), and a resolution of 1 MHz (later 1.4 MHz). The
results are shown in Fig.\,\ref{cofig}, and summarised in
Tables\,\ref{data12co} and \ref{data13co}.

\section{Results }

\subsection{Complex nature of the observed nuclear line profiles}

\nobreak  The complexity of the individual line profiles is
particularly obvious in the lower $J$ transitions of $\co$
(Fig.\,\ref{cofig}). These illustrate the presence of the various
contributions that are due to physically distinct components in the
line of sight to the nucleus (see Israel 1992, 1998). These are: (a)
very broad line emission at $300\,\kms\,<\,V_{LSR}\,<\,800\,\kms$
sampling the rapidly rotating compact CND, fully covered by beams
larger than $20''$, but progressively more resolved in smaller beams;
(b) narrower line emission roughly at
$450\,\kms\,<\,V_{LSR}\,<\,650\,\kms$, most prominent in the lower
$\co$ transitions, that samples the much larger and fully resolved
ETD, and (c) a pattern of absorption against the nuclear continuum
point source (but not the extended dust continuum) in the range
$500\,\kms\,<\,V_{LSR}\,<\,625\,\kms$. Because the continuum emission
from the nucleus is completely unresolved even in the smallest
observing beams, the absorption features sample the molecular ISM in
both the ETD and CND along the line of sight to the nucleus in a very
narrow pencil-beam, less than $0.1''$ across (i.e $<2$ pc).

\nobreak In the remainder of this paper, we will first briefly discuss
the continuum emission, and then concentrate on the {\it line
  emission}, i.e contributions (a) and (b). Analysis and discussion of
the {\it line absorption} (contribution (c)), based on profiles with
higher spectral resolution than shown here, will be presented in a
subsequent paper.

\subsection{Nuclear continuum emission}

\nobreak In the beams used to obtain the velocity-resolved profiles,
the (variable) flux of the nuclear point source is dominant, and dust
continuum emission is negligible (cf Israel \etal 2008).  However, at
the higher frequencies shown in Fig.\,\ref{spirespec}, there is a
steadily increasing contribution by thermal emission from dust
extended over the beam area.  The unresolved nuclear continuum
spectrum can be approximated by a power-law
$F_{\nu}\,\propto\,\nu^{-0.36}$ (Meisenheimer \etal 2007). From {\it
  APEX-1} and {\it APEX-2a} observations of the continuum underlying
the HCO$^{+}$ line emission at 267 GHz and 355 GHz made in June, 2010
(i.e. close to the {\it SPIRE} observing date of August, 2010) we
extrapolate a flux density $F_{461GHz}\,=\,8.2\pm0.3$ Jy. As this is
indistinguishable from the value implied by Fig.\,\ref{spirespec}, we
conclude that at the lowest frequencies observed with {\it SPIRE} the
continuum flux is still wholly due to the nucleus.  At the highest
observed {\it SPIRE} frequencies around 1500 GHz, the nuclear flux has
decreased to $F_{1500GHz}\,\approx\,5$ Jy, and contributes no more
than $15\%$ to the continuum flux measured in the aperture.

\subsection{Spectral line flux distributions }

\subsubsection{Integral CND/ETD line profiles}

\nobreak First, we will analyse the observed (CND/ETD) spectral line
emission without attempting to separate it into the individual ETD and
CND contributions. This allows us to use well-defined integrated line
intensities (such as those from {\it SPIRE}) up to the $J$=(12-11)
$\co$ transition, not hampered by the uncertainties inherent in
component decomposition, and suitable for comparison with measurements
of other galaxy centers. This is particularly important for the
highest observed line transitions, which suffer from increasingly poor
baseline-definition and signal-to-noise ratios in the
velocity-resolved {\it HIFI} measurements. The measured emission line
intensities, and the different beam sizes used, are given in
Tables\,\ref{data12co}, \ref{data13co} and \ref{datacarbon}.

\nobreak In constructing the NGC~5128 CO spectral line ladder we need
to take into account the variation in beam sizes.  The CND and the ETD
have finite but different extent, and the observed emission line
fluxes must be normalised to the same beam-size. To accomplish
this, we have taken the {\it ALMA} Band 6 high-resolution mosaic of
the $J$=2-1 $\co$ emission from Centaurus A obtained in the public
{\it ALMA} Calibration and Science Verification program.  From these
data, we have constructed a series of maps at spatial resolutions
identical to those of our single-dish line observations.  From each of
these maps, we have extracted the central (nuclear) line profile and
determined its integral value. Assuming that the distribution on the
sky of CO emitting clouds is identical for all transitions, we have
determined the beam normalisation factors (BNFs) required to relate
the emission line fluxes to one another. These factors are included
in Table\,\ref{luminosities}.

When more than one independent measurement was available, we took the
error-weighted mean to derive the flux.  In all these cases, the
derived normalised fluxes agree reasonably well with one another, and
also with values obtained from linear interpolation between
measurements with beams larger respectively smaller than $22''$.  For
instance, our normalised flux for the \cii\ line is only $10\%$
above the value interpolated from the HIFI ($11''$ beam) and ISO-LWS
($70''$ beam) measurements published by Unger \etal (2000). This is
quite gratifying, as the large beam normalisation factors appropriate
to such small beams are at the limit of reliability.

\subsubsection{Absorption correction and normalisation} 

Since the absorption occurs against the continuum emission from a
nucleus very much smaller than any of the beams used, the absorption
line intensity, unlike the emission line intensity, is independent of
beam size.  We have determined the magnitude of the absorption by
subtracting the integrated line fluxes from the gaussian-fitted line
fluxes. This amount is once again added to the normalised emission
line flux to finally yield the normalised and corrected line {\it
  luminosities} listed in Table\,\ref{luminosities}. We have adopted
an uncertainty in the absorption line flux of 50$\%$.

Because of the poor definition of the line profiles beyond the $J$=7-6
transition, we have not attempted to correct for absorption losses at
the higher $\co$ transitions. We believe that the effect of this on
the analysis is limited, as Table\,\ref{luminosities} shows that the
magnitude of the flux absorbed against the nuclear continuum decreases
steadily from $J$=2-1 $\co$ upwards. This is probably due to
decreasing absorption optical depths at increasing $J$ levels, and to
the slow flux decrease of the nuclear continuum itself.

For the $\thirco$ transitions we only give the normalised and
corrected luminosities. Because of the relatively low signal-to-noise
ratios of these weaker lines, we have not attempted to process the
individual profiles. Instead, we have divided the $\co$ luminosities
by the $\co/\thirco$ ratios presented in Table\,\ref{data13co}. This
is more accurate because these were obtained by fitting, in each
transition, the $\thirco$ profiles to the much more accurate $\co$
profiles.

\section{Analysis and discussion}
 
\subsection{Comparison of the central CO ladders in different galaxies}

CO line fluxes extracted from {\it SPIRE} spectra have been published
for a limited number of active (mostly Seyfert) galaxies
(Pereira-Santaella, 2013) and also for the relatively nearby NGC~1068
(Spinoglio et al. 2012). {\it SPIRE} data are also available for
star-burst galaxies including the nearby M~82 (Panuzzo \etal 2010) as
well as the more distant and much more energetic Markarian 231 (Van
der Werf \etal 2010), Arp~220 (Rangwala \etal 2011), NGC~6240
(Meijerink \etal 2013), Arp~193 (Papadopoulos \etal 2013).  The latter
are all (ultra)luminous infrared galaxies, or (U)LIRGs). Due to their
much greater distances (D = 42 - 107 Mpc) they were covered in their
entirety, whereas the {\it SPIRE} aperture sampled only the central
region in the nearby galaxies.  Because M~82 is at the same distance
as Centaurus A, the Herschel measurements with {\it SPIRE} (Panuzzo
\etal 2010) and HIFI (Loenen \etal 2010) are directly comparable to
those presented here.  Unlike NGC~5128, M~82 has a star-burst center
lacking an easily identifiable nucleus. In NGC~1068, the {\it
  SPIRE}-SLW aperture covers both the Seyfert AGN and the star-burst,
but the SSW aperture covers only the AGN and its surroundings (cf
Fig. 2 in Spinoglio \etal 2012). It is of interest to compare the {\it
  SPIRE} line spectra of NGC~5128 and these galaxies and identify any
differences.

\begin{table*}
\scriptsize
\caption[]{Mean C/CO line luminosity ratios}
\begin{center}
\begin{tabular}{lcccccc}
\noalign{\smallskip}     
\hline
\noalign{\smallskip}
Ratio &NGC~5128$^{a}$&AGNs$^{b}$&Star-bursts$^{c}$&Nearby$^{d}$&\multicolumn{2}{c}{Milky Way}\\
      &          &          &                  &             & Center$^{e}$& Clouds$^{f}$\\  
\noalign{\smallskip}     
\hline
\noalign{\smallskip}     
CO(10-9)/CO(5-4) &0.26$\pm$0.05&0.38$\pm$0.08&1.25$\pm$0.15& --- & --- & --- \\
\ci(2-1)/\ci(1-0)&2.6$\pm$0.5&1.6$\pm$0.3&3.2$\pm$0.3&  ---      &1.0$\pm0.1$&0.5 - 1.9\\
\ci(1-0)/CO(4-3) &1.2$\pm$0.1&0.6$\pm$0.1&0.4$\pm$0.1&0.6$\pm$0.2&0.6$\pm0.1$&0.05-0.45\\
\ci(2-1)/CO(7-6) &5.7$\pm$0.8&1.6$\pm$0.2&0.7$\pm$0.1&  ---      &1.1$\pm0.2$&0.3$\pm$0.1\\
\noalign{\smallskip}     
\hline
\noalign{\smallskip}
\end{tabular}
\end{center}
Notes: 
(a) {\it SPIRE} data from this paper,
(b) {\it SPIRE} data from Spinoglio \etal (2012); Pereira-Santaella \etal (2013). 
(c) {\it SPIRE} data from Panuzzo \etal (2010); Van der Werf \etal (2010); 
    Rangwala \etal (2011); Meijerink \etal (2013); Papadopoulos \etal (2013); 
    Rosenberg \etal (2013).
(d) Ground-based data from Israel $\&$ Baas (2002); Hitschfeld \etal (2008). 
    Note that this sample overlaps with both the {\it SPIRE} star-burst and {\it SPIRE} 
    AGN samples. 
(e) {\it COBE} data from Fixsen \etal (1999).
(f) From data compiled by Kramer et al. (2004, 2008) 
\label{cratios}
\end{table*}

The right-hand diagram in Fig.\,\ref{spirespec} shows, for each CO
transition, the observed line flux ratio of the luminous star-burst
galaxies and the NGC~5128 center. No errors are indicated, as the
formal extraction errors are very small so that the uncertainties are
dominated by systematic effects. All the low-frequency (SLW) flux
ratios are seen to increase with transition. However, at the highest
(SSW) frequencies, the flux ratios with Arp~220, Arp~193 and the M~82
center are constant, whereas ratios for NGC~6240 (and the ULIRG
Markarian 231 which has an identical CO ladder - see Van der Werf
\etal 2010; Meijerink \etal 2012) keep increasing. This pattern
suggests that the most-highly excited gas fraction in NGC~5128,
emitting in the highest (SSW) transitions, is similar in nature to the
corresponding gas phase in the star-burst galaxies, whereas the bulk
of the molecular gas in NGC~5128 (emitting in the lower transitions of
the SLW region) is much less highly excited than most of the gas in
the star-burst galaxies.

The NGC~5128 CO ladder peaks in the $J$=5-4 $\co$ transition
(corresponding to an upper energy level temperature
$T\,=\,E_{U}/k\,=\,83$ K), and then decreases to low levels barely
reaching a quarter of the peak luminosity in the $J$=10-9 transition
(see, for instance, Fig.\,\ref{pdrmod}, and also
Table\,\ref{cratios}).  The CO emission from all bright-star-burst
galaxy peaks in the $J$=7-6 (155 K) or $J$=8-7 (200 K)
transitions. These galaxies have CO ladders that are relatively flat
between the $J$=5-4 and the $J$=13-12 (500 K) transitions. Their SPIRE
FTS spectra show strong CO lines up to the highest observed
frequencies near 1550 GHz ($J$=13-12), whereas the spectrum of
NGC~5128 exhibits CO lines drowning in the noise beyond $J$=9-8 near
1000 GHz.  In M~82, resembling the (U)LIRGs but more modestly excited,
the CO ladder peaks in the $J$=7-6 transition, and CO line
luminosities decline more rapidly than in the (U)LIRGs but not nearly
as fast as in NGC~5128.  For instance, the $J$=10-9 transition in the
M~82 center still has $60\%$ of the luminosity in the peak transition,
versus $26\%$ for the NGC~5128 center.  The comparison with the
resolved AGN-galaxies is more difficult to make, but except for
NGC~7582 their CO ladders peak in the $J$=4-3/$J$=5-4 transitions,
only slightly below the Cen A peak. All AGN ladders drop significantly
with increasing transition. This behaviour is illustrated by the
CO(10-9) to CO(5-4) luminosity ratios summarized in
Table\,\ref{cratios}.  The lowest ratio (corresponding to the
steepest CO ladder high frequency {\it drop}) is presented by the
NGC~5128 center, followed by the average AGN; the average star-burst
galaxy shows a {\it rise} instead. Thus, the central NGC~5128 CO
ladder is the 'coolest' of all galaxies considered.

\subsection{Remarkably strong [CI] emission lines}

NGC~5128 has a $J$=2-1/$J$=1-0 [CI] luminosity ratio of 2.6
(corresponding to a line brightness temperature ratio of 0.6),
indistinguishable from both M~82 and NGC~1068, and more generally
in-between the {\it mean} values found for the AGN centers and the
star-burst disks (cf. Table\,\ref{cratios}). 

However, {\it both} \ci\ lines in the center of NGC~5128 are rather
bright with respect to the CO lines, unlike the situation in the other
galaxies (Table\,\ref{cratios}).  In NGC~5128, the 809 GHz [CI](2-1)
line is almost six times stronger than its 807 GHz CO(7-6)
neighbour. This very high NGC~5128 [CI]/CO(7-6) ratio is
unparalleled.  The star-burst galaxies show the weakest [CI] lines;
the AGNs have double their relative [CI] line intensity but this still
falls far short of the NGC~5128 value. The 492 GHz [CI] line is also
more intense than the nearby 461 GHz CO(4-3) line. Corresponding
[CI]/CO(4-3) line ratios are much lower in the luminous star-burst
galaxies. They are also lower (but not as much) in the AGNs, the Milky
Way center, and in the 15 nearby galaxies of various type observed
from the ground by Israel $\&$ Baas (2002), and Hitschfeld
\etal. (2008). Among these, high ratios as in NGC~5128 are found only
in NGC~3079, NGC~4826, NGC~4945, M~51, and Circinus. At least three of
the latter galaxies have a nuclear outflow.  Such high ratios provide
a strong hint that an excitation mechanism other than PDR is required
(Israel, 2005), as does the very high [CI]/$\thirco$(2-1) ratio in
NGC~5128. Indeed, the observed [CI] and CO line intensities are
inconsistent with e.g. the PDR models by Kaufman \etal (1999). In
these models, the observed [CI]492GHz line {\it intensity} requires a
density $n_{o}\,\approx\,3000\,\cc$, and a very high radiation field
$G\,\geq\,5\times10^{6}$ G$_{o}$, whereas the intersection of the
[CI]809GHz/[CI]492GHz and [CI]492GHz/CO(1-0) line {\it intensity
  ratios} define a density $n_{o}\,\approx\,1000\,\cc$ and a very low
radiation field $G\,\approx\,10$ G$_{o}$. 

Finally, by comparing the intensities of a particular line obtained
with different apertures, we may obtain an estimate for the size of
the area emitting in that line.  A very compact source will produce
the same brightness in all apertures independent of size, whereas a
very extended source will rapidly become brighter with increasing
aperture size.  We conclude from the CO $J$=4-3, CO $J$=7-6, and [CI]
J=1-0 data in Tables\,\ref{datacarbon} and \ref{data12co} that the
[CI] emission arises from a much more compact region than the $J$=4-3
$\co$ emission but that the sizes of the regions emitting in the
$J$=7-6 $\co$ and [CI] transitions are not very different (implying
that the $J$=7-6 $\co$ is also much more compact than the $J$=4-3
$\co$ distribution).

Thus, the ladder of CO intensities from the (AGN-dominated) central
region of NGC~5128 is quite distinct from the CO ladders representing
extragalactic star-burst environments.  In NGC~5128, the CO ladder
intensities decrease more rapidly and most of the molecular gas is
distinctly less excited than in the other galaxies. Most remarkably,
NGC~5128 has [CI] line emission (much) stronger than that of the
nearest CO lines. This is quite unlike other galaxies, whether normal
or in possession of an AGN or a star-burst, where the [CI] lines are
less prominent.

\begin{figure*}[]
\centering
\includegraphics[width=3.5cm,clip,angle=270]{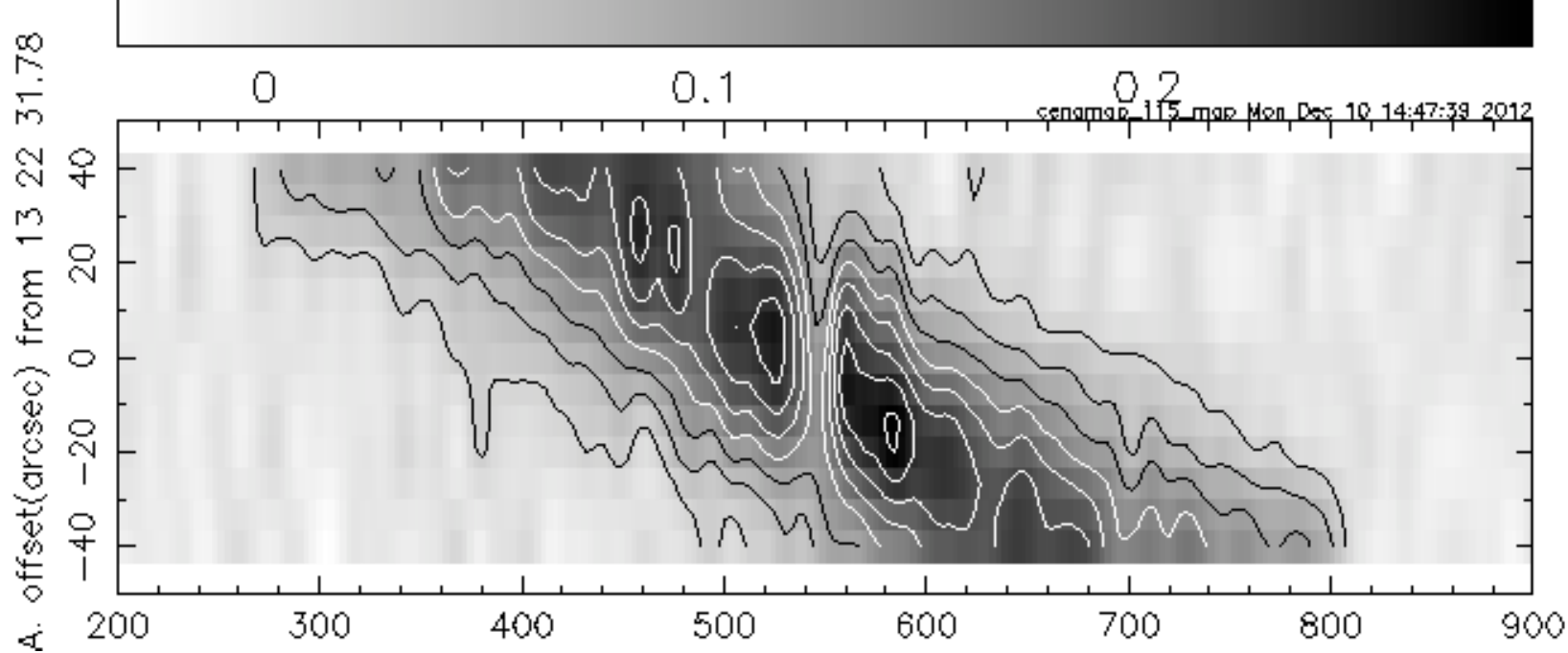}
\includegraphics[width=3.5cm,clip,angle=270]{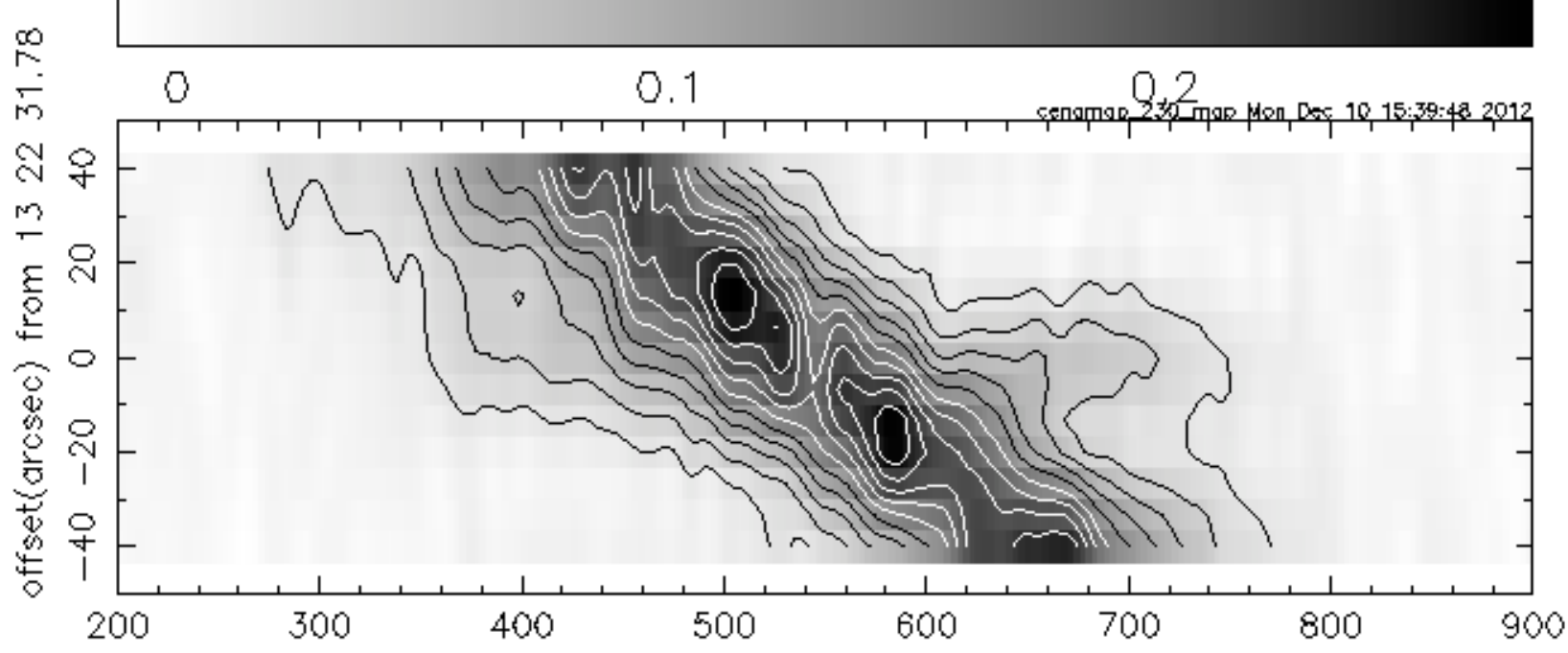}
\includegraphics[width=3.5cm,clip,angle=270]{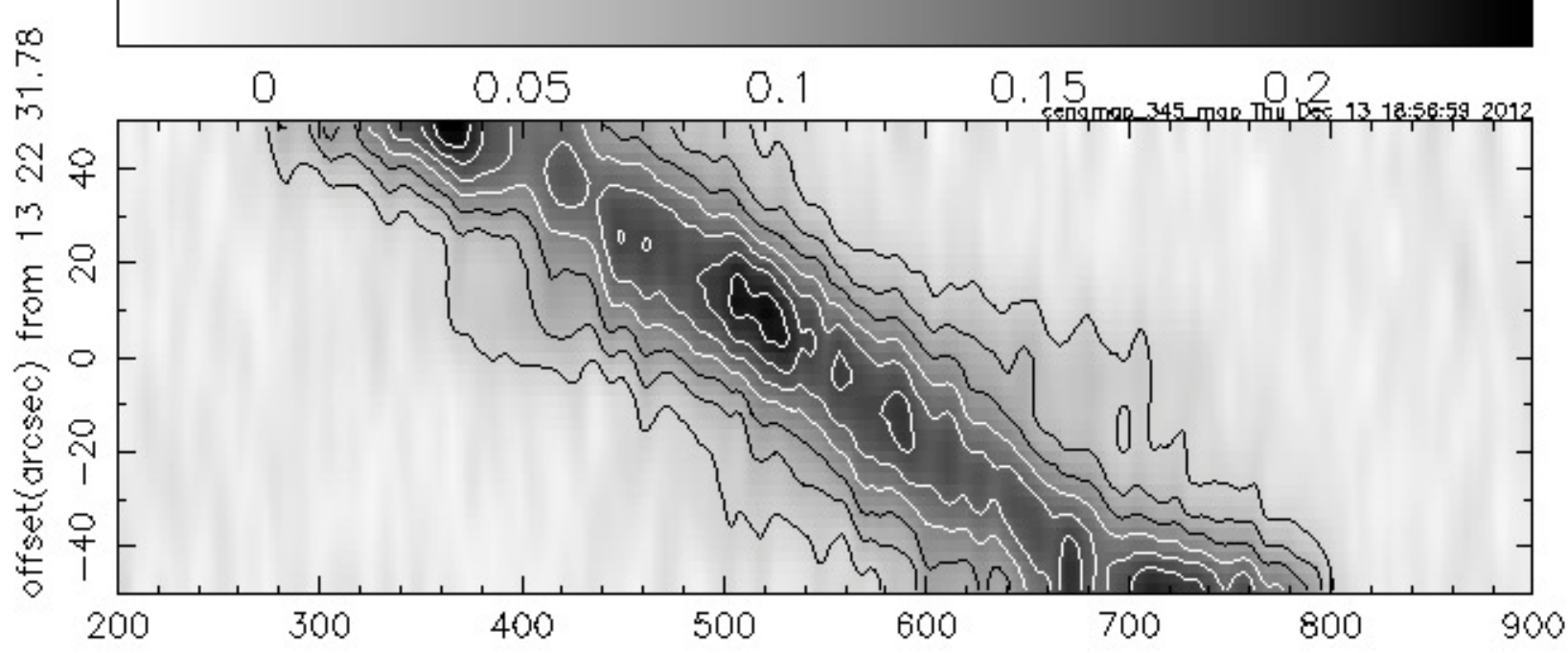}
\includegraphics[width=3.5cm,clip,angle=270]{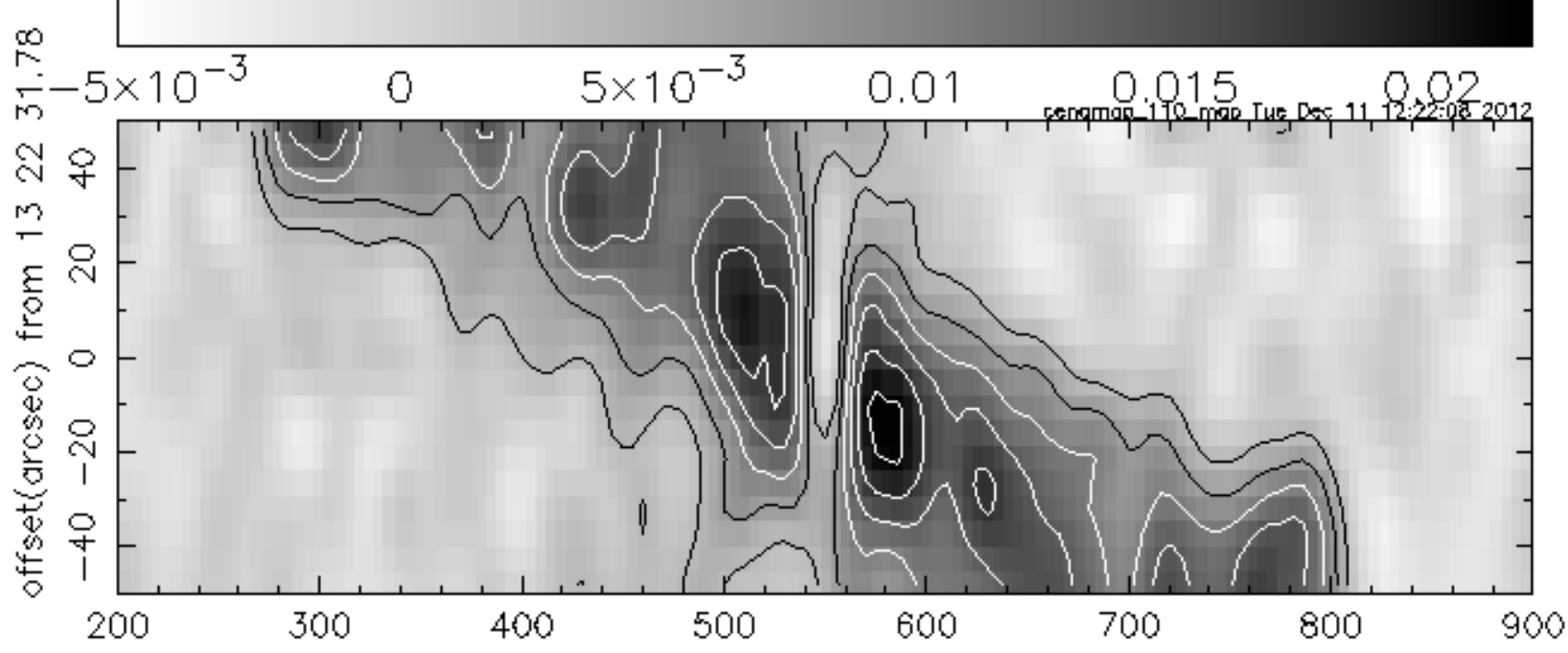}
\includegraphics[width=3.5cm,clip,angle=270]{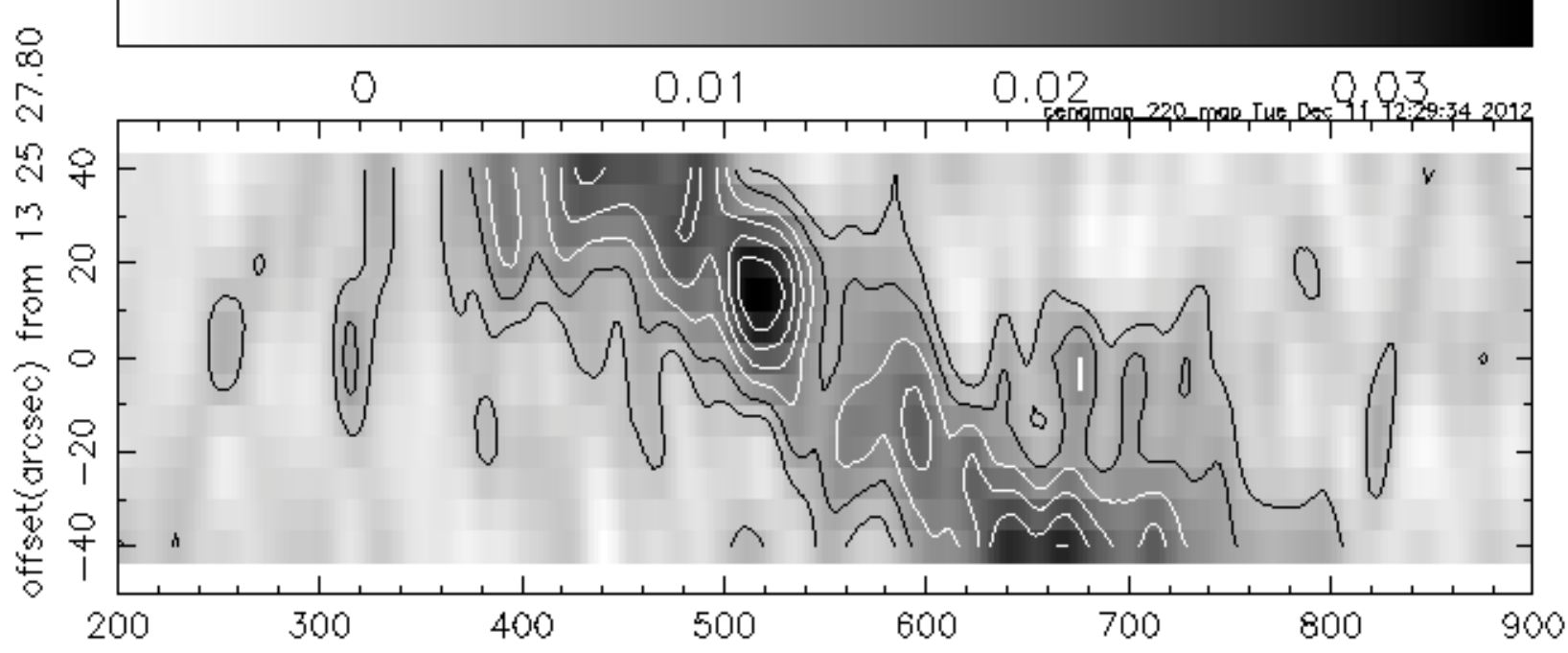}
\includegraphics[width=3.5cm,clip,angle=270]{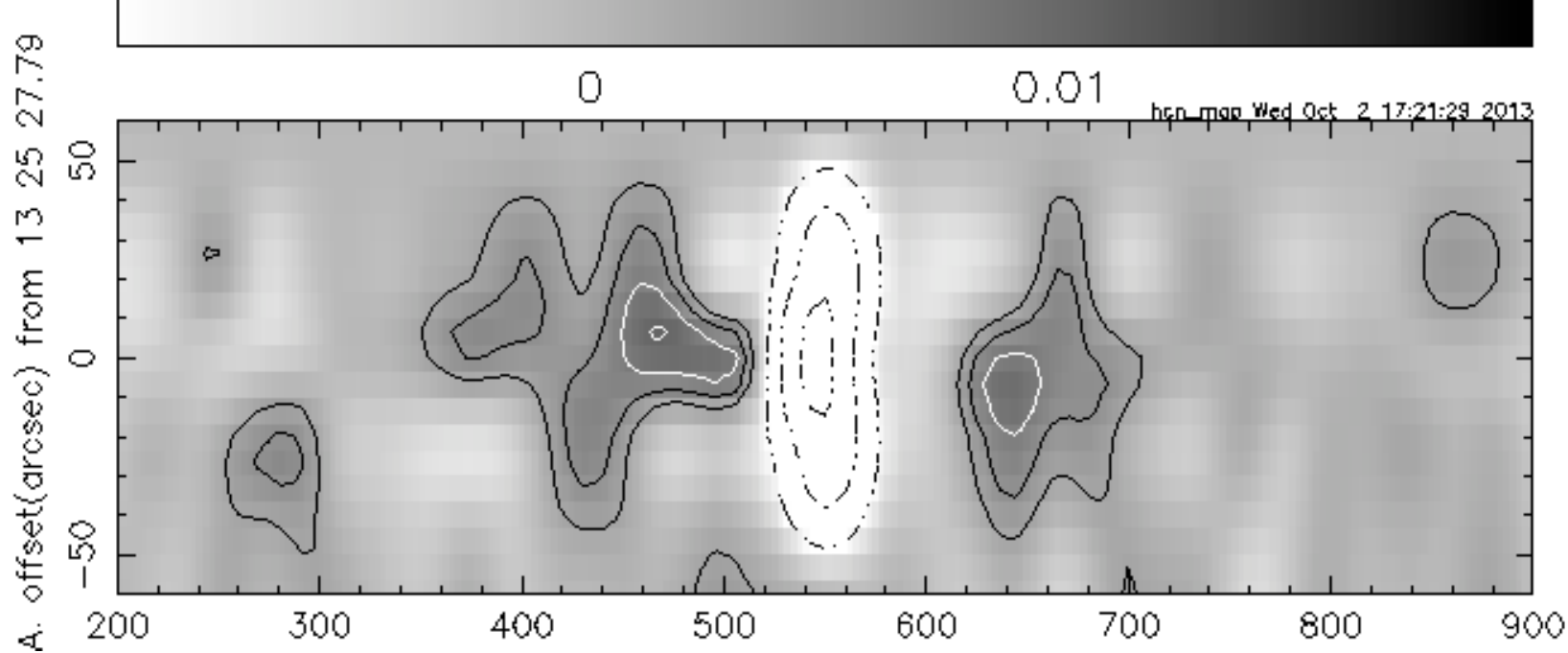}
\caption[]{Position-velocity maps of molecular line emission from the
  central region of Centaurus A, in position angle 125$^{\circ}$
  counter-clockwise from north (data summarised in
  Tables\,\ref{data12co} and \ref{data13co}). Horizontal scales are
  velocity V(LSR) in $\kms$, vertical scales are offsets from the
  nucleus in arcsec. The panels on the left show emission from $\co$;
  the $J$=1-0, $J$=2-1, and $J$=3-2 transitions are shown from top to
  bottom, respectively.  The panels on the right show emission from
  $\thirco$ ($J$=1-0 at the top, $J$=2-1 in the middle). The $J$=1-0
  HCN transition is at bottom right. The $J$=1-0 maps have resolutions
  of $45"-55"$, all other maps have an effective resolution of
  $23''$. In all three $\co$ maps, the contours are at multiples of 50
  mK in main-beam brightness temperature. In the $\thirco$ maps
  contours are at multiples of 5 mK ($J$=1-0) and 10 mK ($J$=2-1). The
  HCN map contours are at multiples of 5 mK. Strong absorption is
  clear in all panels near $V_{LSR}$ = 550 $\kms$. Emission from the
  rapidly rotating compact nuclear disk becomes progressively more
  clear with increasing $J$-transition in the $\co$ panels, and in the
  HCN map.  }
\label{pvmaps}
\end{figure*}

\begin{table*}
\scriptsize
\caption[]{Decomposed spectral line emission distributions}
\begin{center}
\begin{tabular}{lcccccccccc}
\noalign{\smallskip} \hline \noalign{\smallskip} Line &
\multicolumn{4}{c}{$I_{CO}^{a}$} & \multicolumn{6}{c}{$L_{cor}^{b}$} \\
 & \multicolumn{2}{c}{(min. CND fit)} & \multicolumn{2}{c}{(max. CND fit)} 
& \multicolumn{2}{c}{(min. CND fit)} & \multicolumn{2}{c}{(max. CND fit)} 
& \multicolumn{2}{c}{'Best' fit} \\ 
& ETD & CND & ETD & CND & ETD & CND & ETD & CND & ETD & CND \\ 
& \multicolumn{2}{c}{($\kkms$)}& \multicolumn{2}{c}{($\kkms$)} & \multicolumn{2}{c}{($10^{4}$ L$_{\odot}$)} & \multicolumn{2}{c}{($10^{4}$ L$_{\odot}$)} & \multicolumn{2}{c}{($10^{4}$ L$_{\odot}$)}\\  
\noalign{\smallskip} 
\hline 
\noalign{\smallskip}
\multicolumn{9}{c}{The $\co$ ladder} \\ 
\noalign{\smallskip} 
\hline
\noalign{\smallskip} 
$J$=1-0 & 59 & 103 & 18 & 162 & 0.05 & 0.08 & 0.01 & 0.12 & 0.025$\pm$0.015& 0.10$\pm$0.025\\ 
$J$=2-1 & 48 &  45 & 28 &  67 & 0.33 & 0.18 & 0.20 & 0.46 & 0.26$\pm$0.07  & 0.38$\pm$0.15 \\ 
$J$=3-2 & 35 &  28 & 16 &  49 & 0.81 & 0.76 & 0.37 & 1.14 & 0.55$\pm$0.20  & 0.93$\pm$0.20 \\ 
$J$=4-3 & 22 &  15 & 8.4 & 29 & 1.03 & 0.92 & 0.40 & 1.41 & 0.64$\pm$0.30  & 1.14$\pm$0.25 \\ 
$J$=5-4 & 13 &  12 & 5.5 & 32 & 0.90 & 0.87 & 0.39 & 2.23 & 0.59$\pm$0.25  & 1.39$\pm$0.50 \\ 
$J$=6-5 &  4 &  14 & 4.7 & 14 & 0.81 & 1.42 & 0.51 & 1.72 & 0.51$\pm$0.15  & 1.72$\pm$0.20 \\ 
$J$=7-6 & -- &  -- & --  & -- & --   & --   & ---  & ---  & 0.25$\pm$0.05  & 1.6$\pm$0.2   \\ 
$J$=8-7 & -- &  -- & --  & -- & --   & --   & --   & --   & 0.10$\pm$0.05  & 1.5$\pm$0.2   \\
$J$=9-8 & -- &  -- & --  & -- & --   & --   & --   & --   &   ---          & 1.4$\pm$0.15  \\ 
$J$=10-9& -- &  -- & --  & -- & --   & --   & --   & --   &   ---          & 0.8$\pm$0.15  \\ 
$J$=11-10& --&  -- & --  & -- & --   & --   & --   & --   &   ---          & 0.8$\pm$0.25  \\ 
$J$=12-11& --&  -- & --  & -- & --   & --   & --   & --   &   ---          & 0.6$\pm$0.2   \\ 
\noalign{\smallskip} 
\hline 
\noalign{\smallskip}
\multicolumn{9}{c}{The $\thirco$ ladder$^{c}$} \\ 
\noalign{\smallskip}
\hline 
\noalign{\smallskip} 
$J$=1-0 & & & & & 0.006 & 0.018 & 0.002 & 0.028 & 0.003$\pm$0.0015 & 0.010$\pm$0.003 \\ 
$J$=2-1 & & & & & 0.026 & 0.028 & 0.015 & 0.039 & 0.020$\pm$0.005  & 0.033$\pm$0.006 \\
$J$=3-2 & & & & & 0.058 & 0.051 & 0.027 & 0.076 & 0.039$\pm$0.015  & 0.062$\pm$0.013 \\ 
$J$=4-3 & & & & & ---   & ---   & ---   & ---   & --- & --- \\ 
$J$=5-4 & & & & & 0.050 & 0.048 & 0.022 & 0.124 & 0.030$\pm$0.018  & 0.077$\pm$0.035 \\ 
\noalign{\smallskip} 
\hline 
\noalign{\smallskip}
\multicolumn{9}{c}{The carbon $\c$ and $\cp$ lines$^{d}$} \\ 
\noalign{\smallskip} 
\hline 
\noalign{\smallskip} 
\ci (1-0) & & & & & &&&& \multicolumn{2}{c}{ 4.2$\pm$0.8}\\ 
\ci (2-1) & & & & & &&&& \multicolumn{2}{c}{ 11.1$\pm$2.4}\\ 
\cii\ & & & & & &&&& \multicolumn{2}{c}{397$\pm$60 } \\ 
\nii\ & & & & & &&&& \multicolumn{2}{c}{  31$\pm$5 } \\ 
\noalign{\smallskip} 
\hline 
\noalign{\smallskip}
\end{tabular}
\end{center} 
Notes: (a) Emission line intensity normalised to the response of a
$22''$ beam, corrected for absorption line losses. (b) Adopted
emission line luminosity in a $22''$ beam, corrected for absorption
line losses. (c) Using the appropriate $\co/\thirco$ ratios from
Table\,\ref{data13co}. (d) No attempt has been made to decompose
these spectra into ETD and CND contributions, but the ETD 
contribution appears to be weak or even negligible
\label{compladder}
\end{table*}

\subsection{Determination of individual CND and ETD CO ladders}

Notwithstanding the advantages of analysing the combined ETD and CND
line profiles as observed, the fact remains that the ETD and CND are
different features, with potentially different physical properties.
Unfortunately, a unique decomposition of the observed profiles is not
straightforward, especially in case of the high $J$ transitions
observed with {\it Herschel-HIFI} which are relatively noisy, and lack
well-defined baselines. Yet, such a decomposition is required if we
are to analyse the properties of the circumnuclear disk separately
from its surroundings.

To gain a perspective on the CND, we have collected in
Fig.\,\ref{pvmaps} position-velocity (pV) maps of the inner $100''$ of
NGC~5128 in the lower $\co$ and $\thirco$ transitions and in the HCN
$J$=1-0 transition, in position angle P.A. = 125$^{\circ}$
anti-clockwise from north. This is along the heart line of the ETD;
the more compact CND has P.A. $\approx\,145^{\circ}$.  Besides the
data shown above, we have used $J$=1-0 $\thirco$ data from Wild \etal
(1997) and $J$=3-2 $\co$ data from Liszt (2001). The ETD is the
diagonal feature dominating the $\co$ maps, and the CND is the almost
horizontal feature in the map center that is most obvious in the HCN
map.  The absorption lines close to the systemic velocity are (almost)
saturated, and most prominent in the lower-frequency $J$=1-0 HCN,
$\co$, and $\thirco$ maps.  Map line intensities are in temperature
units, in which the (subtracted) continuum emission and the associated
absorption drops with increasing frequency.

The difference between the HCN map on the one hand, and the $\co$ and
$\thirco$ maps on the other hand is striking. The ETD signature
(diagonal feature) is very clear in $\co$ and $\thirco$, but absent in
HCN where only the CND signature (horizontal feature) is very
clear. Towards the ETD. HCN intensities are at least six times weaker
than towards the CND. This factor is a lower limit because of the
strong absorption towards the nucleus affecting the measured line
intensity. Because the critical density for excitation of the HCN
$J$=1-0 line is of the order of $10^{6}\,\cc$, the HCN map implies that
essentially all dense gas in the line of sight towards the Cen A
nucleus is actually concentrated in the CND.

The high CO transitions, from $J$=6-5 onwards, predominantly trace
this same dense molecular gas. The CO $J$=7-6 transition, for
instance, has a critical density $n_{crit}\,=\,4\times10^{5}\,\cc$.
This is comparable to that of the HCN $J$=1-0 emitting gas that we
have shown to be limited to the CND.  Indeed, the line profiles of the
higher $J$ CO transitions in Fig.\,\ref{cofig} clearly show the
signature of the rapidly rotating CND, and provide little or no
evidence for a (narrow-line) contribution from the ETD.  We have also
noted in the preceding section that the $J$=7-6 CO emission is much
more compact than the $J$=4-3 CO emission. We are therefore quite
confident that from the $J$=7-6 transition onwards, the observed CO
line emission is predominantly due to the CND.

The same conclusion cannot be drawn for the lower CO transitions from
$J$=1-0 to $J$=5-4, where the complex nature of the velocity-resolved
profiles in Fig.\,\ref{cofig} clearly implies significant
contributions from both the CND (broad plateau) and the ETD (narrower
peak). We have estimated the relative contributions as follows.

Analysis of the maps in Fig.\,\ref{pvmaps} shows that the profile
widths from the CO in the ETD result from a contribution caused by a
change of rotation velocity across the beam
$\Delta\,V_{rot}\,=\,2.9\pm0.3\,\kms$/arcsec ($156\pm16\,\kms$/kpc)
and an intrinsic contribution with a velocity FWHM of 92$\pm3\,\kms$
which is independent of the observing beam.  This beam-independent
contribution implies the existence of a significant line-of-sight
velocity dispersion $<v_{r}>\,=\,39\pm1\,\kms$ in the ETD.

This result allowed us to decompose the line profiles shown in
Fig\,\ref{cofig} into gaussian components corresponding to the maximum
and the minimum CND contribution respectively.  In both procedures the
ETD contribution was represented by a gaussian with a fixed velocity
half-width corresponding to the relevant beam-size, a central velocity
allowed to deviate from the systemic velocity
($V_{LSR}\,=\,540\,\kms$) by at most 10 $\kms$ to take into account
small pointing errors, and leaving only the amplitude as a completely
free parameter. 

The {\it minimum CND emission (CND$_{min}$)} was found by fitting the
observed profiles with {\it two } additional gaussians with central
velocities fixed at $V_{LSR}\,=\,440\pm10\,\kms$ and
$V_{LSR}\,=\,635\pm15\,\kms$ respectively, and leaving both amplitude
and velocity width as free parameters.  The sum of the two CND
gaussians has a minimum around the systemic velocity, thus maximising
the ETD contribution. In the CND$_{min}$ decomposition, we find
roughly equal amounts of flux for the CND and for the normalized ETD
in all transitions (see Table\,\ref{compladder}).

The {\it maximum CND emission (CND$_{max}$)} was found by fitting the
observed profiles with only {\it one} additional gaussian with free
parameters. This second CND gaussian peaks at roughly the same
(systemic) velocity as the first gaussian representing the ETD
contribution, which thus is minimised. Because the CND is not a
completely filled disk (cf. Israel et al. 1990), a single gaussian
component fitted to the wings of the observed profile will effectively
overestimate the CND contribution. In this decomposition, the ETD
contribution in the normalized beam is only a fifth to a fourth of the
CND luminosity.

Table\,\ref{compladder} summarises the results for both decompositions
($CND_{min}$ and $CND_{max}$) in the CO transitions up to $J$=6-5.
The last two columns contain the mean of these two decompositions
which we believe that best represents the actual situation. The
CND:ETD flux/luminosity ratio in a $22''$ (410 pc) beam is typically
2:1 in the lower transitions. Taking into account the beam filling
factor of the CND, its mean line surface brightness exceeds that of
the ETD by about a factor of four in these transitions.

\subsection{Radiative transfer modelling}

\subsubsection{LVG modelling}

\begin{figure*}[]
\begin{minipage}[t]{18cm}
\resizebox{6.cm}{!}{\rotatebox{0}{\includegraphics*{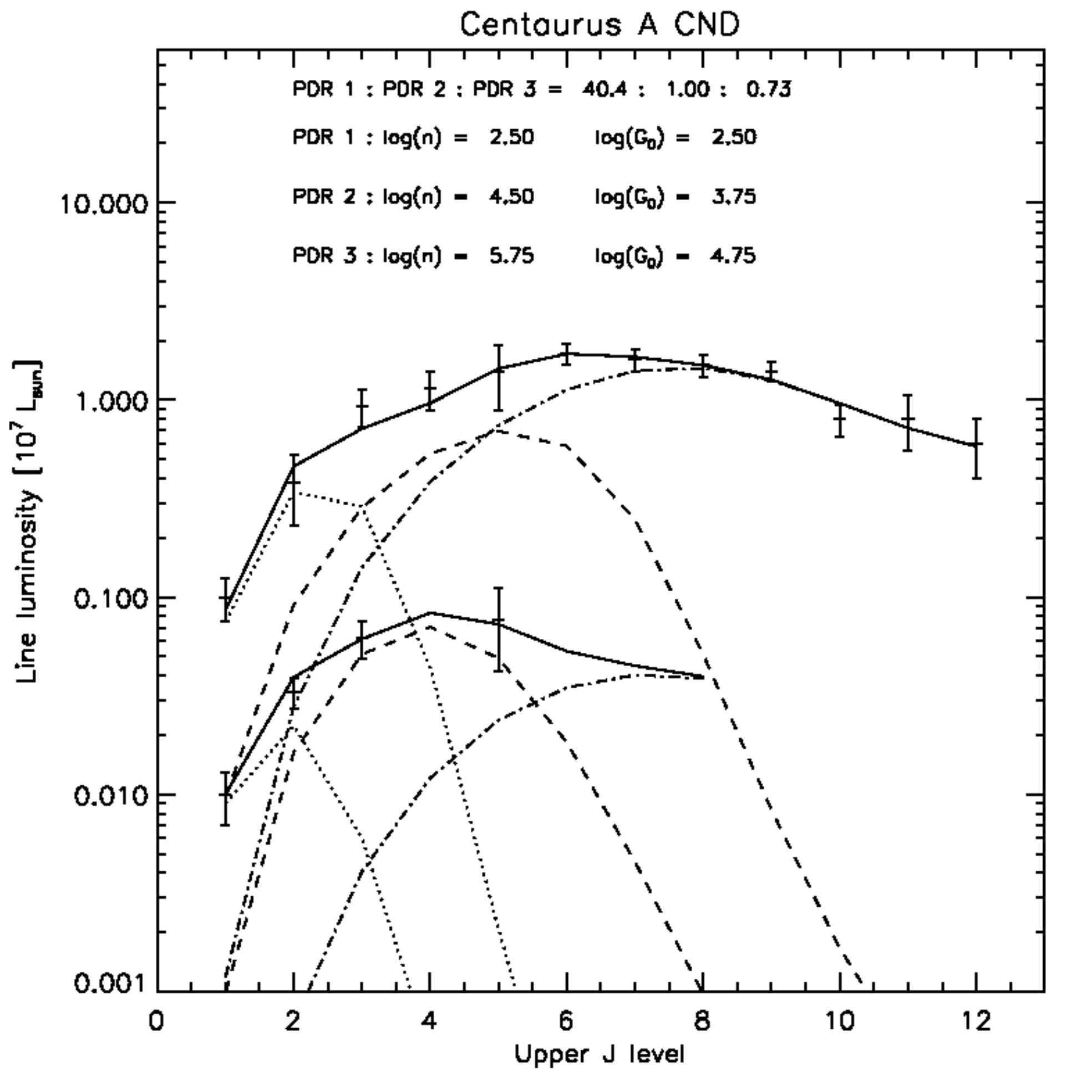}}}
\resizebox{6.cm}{!}{\rotatebox{0}{\includegraphics*{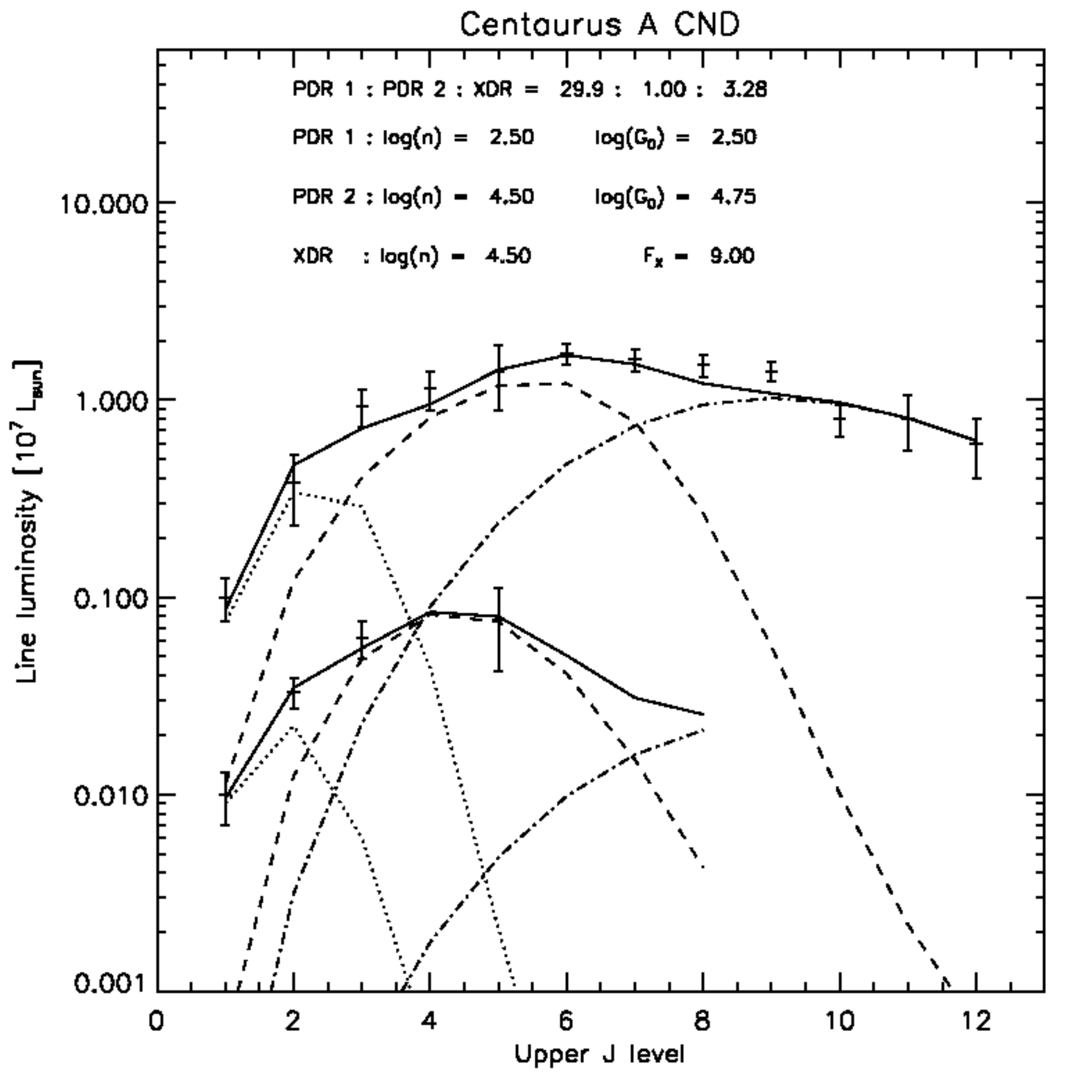}}}
\resizebox{6.cm}{!}{\rotatebox{0}{\includegraphics*{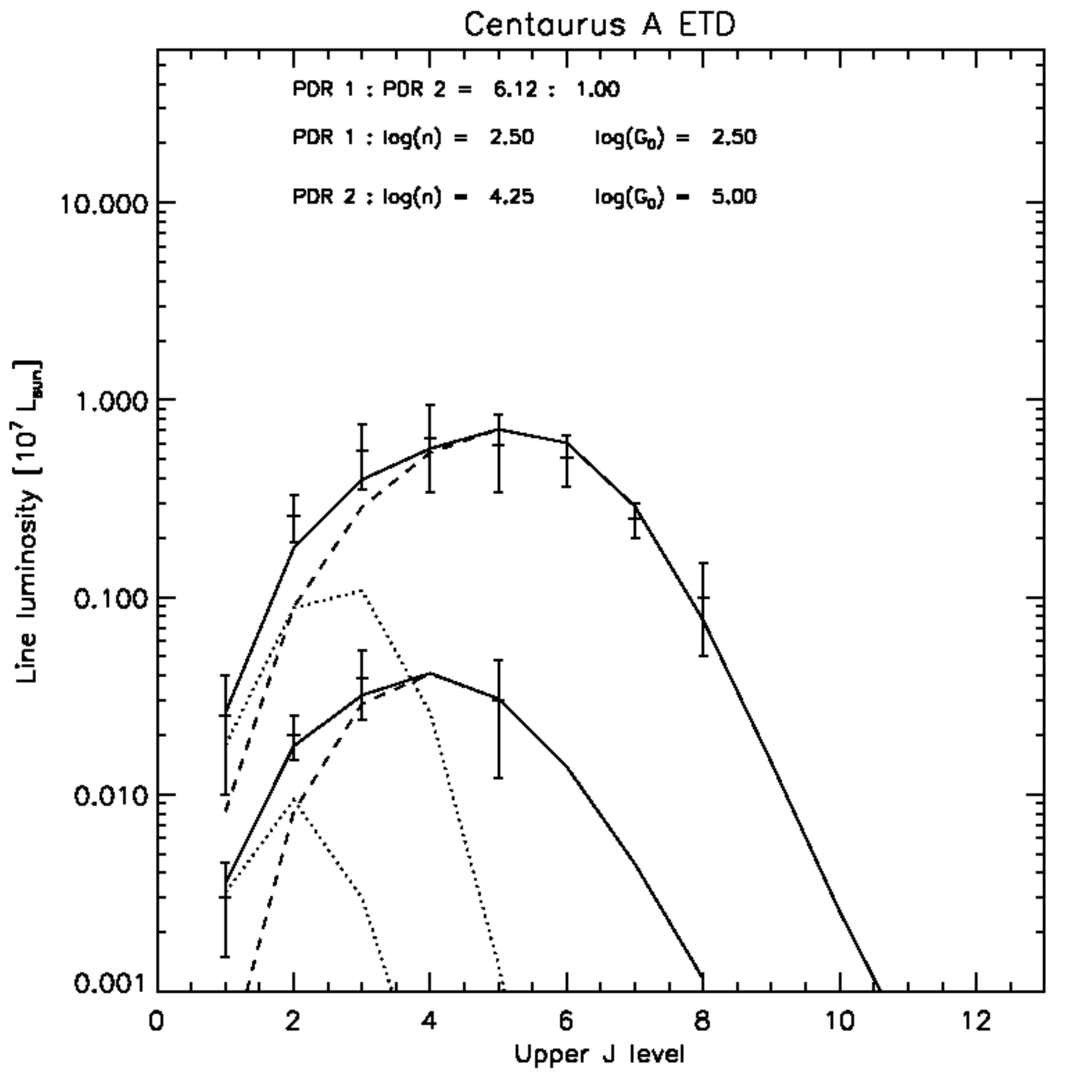}}}
\end{minipage}
\caption[]{Results of the PDR/XDR model fitting to the NGC~5128 CO
  ladders.  Left: Fit to the CND CO ladder with three PDR models;
  center: fit to the CND CO ladder with two PDR models and one XDR
  model; right: fit to the ETD CO ladder with two PDR modelss.}
\label{pdrmod}
\end{figure*}

\nobreak We have modelled the observed $\co$ and $\thirco$ line
intensities and ratios with the {\it RADEX} large velocity gradient
(LVG) radiative transfer models (Jansen 1995; Jansen \etal 1994;
Hogerheijde $\&$ van der Tak 2000;
http://www.strw.leidenuniv.nl/$\,\tilde{}\,$michiel/ratran/ ).  These
codes provide model line intensities as a function of three input
parameters per molecular gas phase: molecular gas kinetic temperature
$T_{\rm k}$, density $n(H_{2})$, and the CO velocity gradient $N({\rm
  CO})$/d$V$.  By comparing model to observed line {\it ratios}, we
may identify the physical parameters best describing the actual
conditions.  We present both the result of modelling the observed
integral CO ladder and those of the derived individual CND and ETD CO
ladders.

\nobreak In the modelling, we assume a constant CO isotopical
abundance [$\co$]/[$\thirco$] = 40 throughout. This value is close to
values found in various galaxy centers (Mauersberger $\&$ Henkel 1993;
Henkel \etal 1993, 1994, 1998; Bayet \etal 2004).  We identify
acceptable fits by searching a grid of model parameter combinations
($T_{\rm k}$ = 10 - 150 K, $n(\h2)$ = $10^{2}$ - $10^{5}\,\cc$, and
$N(CO)/$d$V$ = $6\times10^{15}$ - $3\times10^{18}\,\cm2$) for
(combined) line ratios matching those observed. In the case of
two-phase models, the relative contribution of the two components is
treated as a free parameter.  The physical gas properties undoubtedly
encompass a wider range of temperatures and densities than suggested
by even a two-phase model.  However, lacking a physical model for the
distribution of clouds and their sources of excitation, two phases is
the maximum that can be considered fruitfully, especially as only the
well-defined lower $J$ CO transitions have $\thirco$ observations that
allow us to break the $T_{k}-n(\h2)$ degeneracy inherent to the $\co$
ladder.

\nobreak We find that no combination of parameters from a single-phase
gas provides a satisfactory fit to the observed line intensities. This
is not unexpected in view of the profile complexity involving the two
distinct contributions from ETD and CND. Consequently, we have also
modelled the $\co$ and $\thirco$ simultaneously with two molecular gas
phases.  Although the available $\thirco$ measurements break much of
the strong temperature-density degeneracy inherent to $\co$
measurements, we still find three different possible solutions.  

The first set of solutions puts all gas at an elevated temperature of
$T_{kin}$ = 150 K. In this set, two thirds of the emission comes from
a low-density first phase ($n(\h2)\,=\,100\,\cc$), and one third from
a moderate-density second phase ($n(\h2)\,=\,1000\,\cc$).

In the second set of possible solutions, the two phases have similar
modest densities and elevated temperatures. They differ mainly in
their CO velocity gradient ($3\,\times\,10^{16}\,\cm2\,/\kms$ versus
$4.5\,\times\,10^{17}\,\cm2\,/\kms$, for the second phase). About
$80-90\%$ of the emission arises from the first phase with kinetic
temperatures of 100-150 K, and densities that could be as low as 100
$\cc$ or as high as 1000 $\cc$, but are most likely around 500
$\cc$. The remaining $20-10\%$ arises a slightly warmer, and slightly
more tenuous second phase ($T_{\rm kin}\,\approx\,150$ K,
$n(\h2)\,=\,100-500 \cc$).

The final set of solutions combines an again very similar first phase
($T_{kin}$ = 100 K, $n(\h2)$ = 500 $\cc$, with a much denser and
colder second phase ($T_{kin}$ = 20-30 K,  $n(\h2)\,=\,10^{5}
\cc$). 

We emphasize that the different physical environments represented by
the solutions are observationally indistinguishable. Moreover, they
are not mutually exclusive but may apply simultaneously and sample a
multi-phase structure of the ISM: very dense, cold clumps embedded in
a more tenuous warm molecular gas. 

Thus, the LVG analysis of the integral line profile requires the
presence of significant amounts of warm gas at temperatures of 100-150
K at moderate densities of typically $500\,\cc$, but does not
distinguish between the situation where this is all, where two thirds
of the emission is from a {\it more tenuous} warm gas ($100\,\cc$), or
where $10-15\,\%$ of the emission comes from a {\it much more dense,
  cold} gas. As it turns out, only the latter case fulfils the additional
constraints posed by considering the circumnuclear disk and the
extended thin disk separately.

When we fit LVG models to the individual CND and ETD components, once
again no single phase model provides a reasonable fit to the observed
CO line ladders.  However, the CND CO ladder is well-fitted by two
gas phase components in a narrow range, involving a dense
($n_{\h2}\,=\,10^{4}\,\cc$) and a much more tenuous
($n_{\h2}\,=\,300\pm200\,\cc$) gas. The properties of the dense gas
are well-defined, with a low temperature $T_{kin}\,=25\pm5$ The
temperature of the tenuous gas temperature is not so tightly
constrained. It may be at $T_{kin}\,=60-100$ K, in which case it will
be responsible for three quarters of the CO emission in the $J$=2-1
transition, but it may also be at a much lower $T_{kin}\,=20-30$ K in
which case its $J$=2-1 emission share will be about $50-60\%$.

The physical parameters of the ETD gas are better determined and
differ markedly from those of the CND gas.  Most ($65\pm15\%$) of the
emission in the $J$=2-1 transition) is from a gas with a
well-established high density ($n_{\h2}\,=\,10^{4}\,\cc$) and low
temperature $T_{kin}\,=\,30$ K.  The remainder of the gas (responsible
for $35\pm15\%$ of the emission) has a very low density
($n_{\h2}\,=\,100\,\cc$) at an elevated temperature
$T_{kin}\,=\,100\pm50$ K.

\begin{table}
\scriptsize
\caption[]{CND model physical parameters}
\begin{center}
\begin{tabular}{lcccc}
\noalign{\smallskip} 
\hline 
\noalign{\smallskip} 
             & \multicolumn{2}{c}{LVG model} & \multicolumn{2}{c}{PDR model} \\
                   & CND   &   ETD     &  CND         &    ETD \\
\noalign{\smallskip} 
\hline 
\noalign{\smallskip}  
\multicolumn{5}{c}{ISM phase 1}\\
\noalign{\smallskip} 
\hline 
\noalign{\smallskip}  
density $n(\h2)\,(\cc$)               &    300 &  100   &  300  & 300 \\  
temperature $T_{k}$ (K)               &  25-80 &   100  &  ---  & ---  \\
UV radiation field $G\,(G_{0})$       &    --- &  ---   &  300  & 300  \\
Column density $N_{H}\,(10^{21}\cm2)$  &  5-10  &        &  13    & \\
\noalign{\smallskip} 
\hline 
\noalign{\smallskip}  
\multicolumn{5}{c}{ISM phase 2}\\
\noalign{\smallskip} 
\hline 
\noalign{\smallskip} 
density $n(\h2)\,(\cc$)              & 10000  & 10000  &  30000 & 18000 \\  
temperature $T_{k}$ (K)              &    25   &  30   &    ---  & ---  \\
UV radiation field $G\,(G_{0})$      &   ---   & ---   & 3000-30000& 10000\\
Column density $N_{H}\,(10^{21}\cm2)$ &  2-4   &       &    13  & \\
\noalign{\smallskip} 
\hline 
\noalign{\smallskip}  
\multicolumn{5}{c}{either ISM phase 3: PDR}\\
\noalign{\smallskip} 
\hline 
\noalign{\smallskip}  
density $n(\h2)\,(\cc$)               &   ---  & ---   & 550000 & \\  
UV radiation field $G\,(G_{0})$       &   ---  & ---   &  55000 &  \\
Column density $N_{H}\,(10^{21}\cm2)$  &   ---  & ---   &   7    & \\
\noalign{\smallskip} 
\hline 
\noalign{\smallskip} 
\multicolumn{5}{c}{or ISM phase 3: XDR}\\
\noalign{\smallskip} 
\hline 
\noalign{\smallskip}  
density $n(\h2)\,(\cc$)                  &  --- & ---  & 30000 & \\  
X-ray flux $F_{x}\,({\rm erg}\,\cm2\,{\rm s}^{-1})$ & --- & --- & 5-16 & \\
Column density $N_{H}\,(10^{21}\cm2)$    &  --- & --- & 10-100 &   \\
\noalign{\smallskip} 
\hline 
\noalign{\smallskip}  
\end{tabular}
\end{center} 
\label{CNDmodel}
\end{table}

\subsubsection{PDR/XDR modelling}

In order to further investigate the physical conditions and excitation
mechanisms, we have also applied the PDR/XDR models by Meijerink $\&$
Spaans (2005) and Meijerink et al. (2007). In XDRs, the excitation is
dominated by X-ray photo-ionization heating, i.e., the Coulomb
interaction of keV electrons with thermal electrons. The heating
efficiency of X-rays is of the order of 10 to 40 percent, much more
efficient than the photo-electric heating ($\sim 0.3$ to
$1.0$~percent) in PDRs. The ionisation is driven by primary and
secondary X-ray ionisations, and the ionisation fraction can be larger
than $x_e \sim 0.1$, three orders of magnitude higher than in
PDRs. The high degree of ionisation is able to drive an active
ion-molecule chemistry and makes it possible to maintain high
abundance levels of molecules at high temperatures, $T > 300$~K.
This expresses itself in much larger column densities of warm gas in
XDRs, which betray themselves by much more intense higher $J$ CO
transitions.

The PDR/XDR code requires three input parameters: a density $n$, a
surface area covering factor (in this case relative to that of
component PDR2), and an incident UV flux in units of the
one-dimensional Habing (1968) field $G_{0}$ ($=\,1.6 \times10^{-3}$
erg $\cm2$ s$^{\-1}$), or an X-ray energy flux $F_{X}$ (in units of
$\cm2$ s$^{\-1}$), respectively. We constrained the PDR parameters to
be fitted by using the relatively well-established gas volume
densities supplied by the LVG modelling for the first two gas
components. The resulting fits are shown in Fig.\,\ref{pdrmod}. The
boxes on the left and in the center show fits to the CO ladder of the
CND using three PDRs, and two PDRs and one XDR, respectively. The box
on the right e shows the fit to the CO ladder of the ETD in the same
line of sight, which requires only two PDRs. Throughout, the major
contributor to the low-$J$ emission of both $\co$ and $\thirco$ (PDR1)
is well-represented by the selected density $n\,=\,300\cc$ and a
moderate incident radiation field $G\,=\,300\,G_{0}$. It has a large
surface filling factor compared to PDR2, the component dominating the
mid-$J$ transitions ($3\leq$J$<8$).  PDR2 requires a more intense
incident radiation field $G\,=\,10^{3.75}\,-\,10^{4.75}\,G_{0}$ with
higher densities $n\,=\,3\times10^{4}\,cc$ for the CND and
$n\,=\,10^{4}\,cc$ for the ETD. Its relatively small surface filling
factor implies that the dense, strongly irradiated component exists
mostly or entirely in clumped form. We note that the fits are not
unique, and the ones presented here are merely those with the lowest
chi-square values. Especially the radiation field is not
well-constrained. The CO emission emitted by PDRs is not very
sensitive to the radiation field, as it originates mostly from the
UV-shielded part of the clouds.

The third component is a significant contributor only in the high $J$
transitions ($J\geq8$). Its existence is required by the CND CO ladder
which reliably extends to the higher $J$ transitions (Section 4.3).
This high-energy component represents either extreme PDR conditions
(PDR3), or X-ray irradiation (XDR). The extreme PDR case must have
very high densities of the order of $n\,=\,10^{6}\,\cc$ (consistent
with the detection of HCN emission from the CND) and radiation fields
$G\,=\,55 000\,G_{0}$. It has a slightly smaller filling factor than
that of PDR2, and also has a three times lower total column density,
$N_{\rm H}\sim 7.3 \times 10^{-21}$~cm$^{-2}$. On the other hand, if
it represents an XDR, the density does not need to exceed that of PDR2
($n\,=\,10^{4.5}\,\cc$), with an X-ray radiation field of
$F_{\rm}\,=\,9.0 $~erg~cm$^{\-2}$~s$^{\-1}$ and a column density of
$N_{\rm H} = 3\times 10^{22}$~cm$^{-2}$. In that case, its surface
filling factor is a few times than that of PDR2 (but still much less
than that of PDR1). Equally good fits are, however, obtained for XDR
phases with column densities ranging from $N_{\rm H} = 10^{22}$ to
$10^{23}$~cm$^{-2}$, and concomitant X-ray fluxes of $F_{\rm X}$ of 5
and 16 erg~cm$^{-2}$~s$^{-1}$. respectively.  As
Fig.\,\ref{compladder} shows, the observational data do not
significantly constrain the CND physics because both PDR3 and (a
variety of) XDR models provide comparably good fits to the CO ladder.
Observationally, the appropriate models can only be identified by CO
line intensities at e.g. the $J$=15-14 transition and beyond which are
presently lacking.

\subsection{Beam-averaged molecular gas properties}

In order to derive overall molecular gas column densities from the LVG
data, we assume that only about a quarter of all carbon is in the
gas-phase ($\delta_{C}=0.27$), the remainder being tied down in dust
grains. In a forthcoming paper, we find that the ISM of both the CND
and the ETD is characterized by a metalicity of about 0.7-0.8 times
that of the Solar Neighbourhood.  We assume a [C]/[H] elemental
abundance $x_{c}\,=\,1.6\times10^{-4}$ which leads us to expect expect
a neutral gas-phase $N_{\rm C}/N_{\rm H}=4.5\times10^{-5}$.  The
chemical models by van Dishoeck $\&$ Black (1988), updated by Visser,
Van Dishoeck, $\&$ Black (2009) show a strong dependence of the ratio
of atomic carbon to molecular carbon monoxide column densities on the
total carbon $N_{\rm C}=N(\rm C)+N(\rm CO)$ and molecular hydrogen
$N(\h2)$ column densities.  Thus, each value of $N(\h2)$ is associated
with a unique value of $N(\rm C)/N(\rm CO)$ {\it and} a unique value
of $N_{\rm C}=N(\rm C)+N(\rm CO)$ which can be derived from the
models.

\begin{table}
\scriptsize
\caption[]{NGC~5128 CND properties}
\begin{center}
\begin{tabular}{lc}
\noalign{\smallskip} 
\hline 
\noalign{\smallskip}  
Projected dimensions (pc)                  &  390 x 195 \\  
Velocity width FWHM $\Delta V\,(\kms)$     &  375 \\        
Surface area filling factor                &  0.10$\pm$0.02$^{a}$ \\
\noalign{\smallskip} 
\hline 
\noalign{\smallskip}
Mean column density$^{a}\,N_{H}\,(\cm2)$     &  $(0.3-1.5)\times10^{23}$\\ 
Gas mass$^{b,c}\,M_{gas}\,({\rm M}_{\odot})$    &  $ 8.4\times10^{7}$ \\
Conversion factor$^{d}\,X\,(\cm2/\kkms)$    &  $ 4\times10^{20}$ \\
\noalign{\smallskip} 
\hline 
\noalign{\smallskip}
CO luminosity$^{e}\,L_{CO}\,(\rm L\_{\odot})$ & $(3.6-5.0)\times10^{5}$ \\ 
C$^{o}$ luminosity $L_{CI}\,(\rm L\_{\odot}$  & $(1.5\pm0.3)\times10^{5}$\\
C$^{+}$ luminosity $L_{CII}\,(\rm L\_{\odot}$ & $(4.0\pm0.6)\times10^{6}$\\ 
\noalign{\smallskip} 
\hline 
\noalign{\smallskip}
\end{tabular}
\end{center} 
Notes: (a) LVG two-phase and PDR three-phase result; uncertainty $50\%$. 
(b) Uncertainty factor of three; lower limit from LVG assuming all H in $\h2$, 
upper limit from PDR/XDR modelling assuming no mechanical heating in PDR1.
(c) Assuming a $35\%$ mass contribution by helium. 
(d) Uncertainty factor of two.
(e) Sum of all CO transitions up to $J$=15-14
(\label{CNDprop}
\end{table}

From the LVG analysis, we find very similar beam-averaged column
densities for the CND ($N({\rm CO})=0.34\times10^{18}$) and for the ETD
($N({\rm CO})=0.29\times10^{18}$ $\cm2$).  We also find that in both the
CND and the ETD about two to three times more carbon is in atomic than
in molecular form, and we obtain a beam-averaged total hydrogen column
$N_{\rm H}=(3.4\pm0.1)\times10^{22}\,\cm2$ for the CND, and $N_{\rm
  H}=(1.0\pm0.1)\times10^{22}\,\cm2$ for the ETD.  Taking into account
that the ETD fills all of the normalized beam, and the CND only half
of it, we derive a four times higher molecular gas surface-filling
factor ($0.10\pm0.01$) for the CND.

The molecular gas column density is found by subtracting the neutral
hydrogen \hi\ from the total hydrogen column density.  Measurements of
the \hi\ {\it emission} suggest its contribution is quite small in the
central few hundred parsecs (see Struve et al. 2010). On the other
hand, \hi\ {\it absorption} line measurements imply very high \hi\
column densities (typically $10^{22}-10^{23}\,\cm2$), but these apply
to a pencil-beam area of order $10^{5}$ smaller and are related to
material much less extended than the scale of the CND. Neglecting a
possible (but probably small) contribution by atomic hydrogen, we find
for the CND a beam-averaged molecular hydrogen column density
$N(\h2)=(1.7\pm0.1)\times10^{22}\,\cm2$.  The corresponding mass is
$M_{gas}\,=\,4.7\pm0.5\,10^{7}\,\Msun$, including a $35\%$ helium
contribution.  In the ETD the \hi\ column may not be negligible.  For
$N(HI)\,\leq\,0.15\times10^{22}\,\cm2$ (cf Struve \etal 2010) we
obtain for the ETD $N(H_{2} = 0.9\times10^{22}\,\cm2$.

The PDR model analysis produces higher gas masses. The pure PDR-model
yields a total gas mass $M_{gas}\,=\,1.5\times10^{8}\,\Msun$ which
includes a $35\%$ contribution of helium. A fraction of 0.95 of this
mass resides in the lowest-excitation phase (PDR1), whereas the
densest, highest-excitation gas (PDR3) mass is of the order of a per
cent of the total.  If we assume that the highest observed CO $J$
transitions come from gas excited by X-rays rather than by UV-photons,
the mass of this phase can be substantially higher. For instance, if
we assume an XDR density identical the the PDR2 gas, we obtain a
(helium-corrected) gas mass $M_{gas}\,=\,1.7\times10^{8}\,\Msun$, of
which $15\%$ is in the X-ray irradiated phase. This {\it total} gas
mass is only slightly higher than that of the pure PDR case.  In the
fraction of the ETD measured by our beam, the mass density is no more
than a quarter of that of the CND, and only $15\%$ is in a denser,
more excited phase (PDR2). As shown in Fig.\,\ref{pdrmod}, there is no
third phase of very dense gas highly excited by X-rays or their
equivalent. In the pure PDR case, the mass of dense gas in the CND is
not much higher than that in the ETD-beam. However, in the XDR case
the mass of dense gas is ten times higher in the CND than in the
ETD-beam. This is in better agreement with the the observation that
the HCN emission (Table\,\ref{data13co} and Fig.\ref{pvmaps} traces
dense gas almost exclusively in the CND.

In the PDR/XDR models the atomic gas mass fraction ranges between
one-third to half of the total gas mass. This is because quite large
radiation fields are needed to obtain PDR excitation conditions that
are able to reproduce the observed line fluxes. This can, however, be
mitigated by including a small amount of mechanical heating
(cf. Kazandjian et al. 2012; Kazandjian et al. 2013 submitted). When
sloshing motions of the gas are responsible for part of the total
heating, especially in the regions where the gas is shielded from UV
radiation, the incident UV fluxes can be substantially lowered (yielding
a smaller atomic hydrogen fraction), while obtaining the same CO model
fluxes. 

The physical properties of the NGC~5128/Centaurus A circumnuclear disk
are summarised in Table\,\ref{CNDprop}. For the best estimate of the
CND mass we use the geometric mean of the LVG and PDR results. This
brings the mass to $M_{gas}\,=\,8.4\times10^{7}\,\Msun$ with a
corresponding $N(\h2)$-to-$I_{\co}$ ratio $X$ = 4$\times10^{20}$
$\cm2/\kkms$, about twice the 'standard' Milky Way ratio.  We cannot
determine a useful mass value for the small fraction of the ETD
included in our beams, but from our analysis we estimate an $X$-value
not very different, and possibly a bit closer to that of the Milky
Way.  This is the first time the CND gas mass has been determined
rather than guessed. For instance, a much lower CND mass was given by
Espada \etal (2009), but this was based entirely on an assumed very
low $X$-factor of 0.4 $X_{GAL}$. This appears to be appropriate for
'normal' AGNs, but does not adequately describe the Cen~A situation.

In Table\,\ref{CNDprop} we also compare the integrated CO luminosity
(using the models to extrapolate the observed CO line luminosities up
to the $J$=15-14 transition) with the observed neutral and ionized
carbon luminosities. In case of significant X-ray excitation, the
total CO luminosity may be somewhat higher than given here, because of
the contribution of the very high-$J$ transitions. Nevertheless, from
the comparison in Table\,\ref{CNDprop} we may conclude that neutral
carbon cooling of the CND is at most half of that provided by carbon
monoxide. In contrast, the cooling by ionized carbon is about an order
of magnitude higher than the CO cooling.

\section{Conclusions}

\begin{enumerate} 
\item We have measured $\co$ fluxes from the central region of
  NGC~5128 (Centaurus A) in transitions up to $J$=12-11, as well as
  the fluxes from the submillimeter \ci and \cii lines. In
  addition, we have presented high-S/N velocity-resolved profiles of
  $\co$ emission in the $J$=1-0 through $J$=5-4 and of $\thirco$
  emission in the $J$=1-0 through $J$=3-2 transitions.
\item $\co$ line luminosities, normalized to a $22''$ beam (410 pc)
  increase up to $J$=5-4, and then steadily drop to low values. This
  drop is more pronounced than that seen in Seyfert AGNs, and much
  more pronounced than seen in star-burst galaxies.
\item Both \ci\ lines are more luminous than the adjacent ($J$=4-3,
  and $J$=7-6) $\co$ lines. This behaviour is not seen in other
  (star-burst or AGN) galaxies, and is thus unique to the Cen~A
  center. The \ci\ line intensities and their ratio to CO line
  intensities are inconsistent with standard PDR model predictions.
\item Detailed analysis shows that up to $J$=6-5, about a third of the
  $\co$ flux observed in the normalised $22''$ beam is contributed by
  the extended thin disk (ETD) embedded in NGC~5128, and unrelated to
  the compact circumnuclear disk (CND) just contained within the
  normalised beam. The CND contains a high proportion of very dense
  gas, in contrast to the ETD. At transitions above $J$=6-5, the $\co$
  emission is completely dominated by the CND contribution.
\item We have decomposed the observed CO spectral line ladder into
  individual (total) CND and (representative) ETD CO ladders. The ETD
  ladder peaks in the $J$=4-3/$J$=5-4 transition and then drops
  rapidly.  The CND ladder peaks in the $J$=6-5 and $J$=7-6
  transitions before dropping more slowly.
\item LVG and PDR/XDR model analysis of the CND $\co$ fluxes shows that
  most of the molecular gas mass resides in a relatively cool
  ($T_{kin}$ = 25 - 80 K), not very dense ($n_{\h2}\,\approx\,300\,\cc$
  gas phase if the CND is heated exclusively by UV photons (PDR). 
\item A small fraction of the gas in the CND is more highly excited
  and has much higher densities (typically $3\times10^{4}\,\cc$).
\item In the CND but not in the ETD a third, more highly excited,
  high-density phase must also be present, either in the form of an
  extreme PDR or in the form of an XDR. 
\item The CND has a total gas mass
  $M_{CND}\,=\,8.4\times10^{7}\,\Msun$ (uncertain by a factor of two)
  which is about $10\%$ of the mass of the much larger ETD.
\item The CO-$\h2$ conversion factor ($X_{CND}$ is
  $4\times10^{20}\,\kkms$ (uncertain by a factor of two), about
  twice the local Milky Way factor $X_{SN}$.
\end{enumerate}

\begin{acknowledgements}
  We thank Markus Schmalzl for help with the {\it ALMA SV} data. We
  also thank all facility observers, engineers, and support scientists
  who by their often anonymous efforts have made possible the
  collection of data presented in this paper.
\end{acknowledgements}

\end{document}